\documentclass[journal]{IEEEtran}

\usepackage{cite}
\usepackage{textcomp}
\usepackage[utf8]{inputenc}
\def\BibTeX{{\rm B\kern-.05em{\sc i\kern-.025em b}\kern-.08em
    T\kern-.1667em\lower.7ex\hbox{E}\kern-.125emX}}

\usepackage{tabularx}
\usepackage{bigstrut}
\usepackage{multirow}
\usepackage{bm}
\usepackage{graphicx}

\usepackage{xcolor}
\usepackage{tikz}
\usepackage{pgfplots}

\usetikzlibrary{pgfplots.groupplots, pgfplots.fillbetween, decorations.pathreplacing,calc}
\usetikzlibrary{arrows.meta}
\usetikzlibrary{fit}
\pgfplotsset{compat=1.3}

\usepackage[ruled]{algorithm2e}
\SetKwFor{For}{for\hspace*{0.3mm}(\hspace*{-0.6mm}}{\hspace*{-0.6mm})}{}

\SetCommentSty{myalgofont}
\makeatletter
\newcommand{\nosemic}{\renewcommand{\@endalgocfline}{\relax}}
\newcommand{\dosemic}{\renewcommand{\@endalgocfline}{\algocf@endline}}
\makeatother

\usepackage[para]{threeparttable}

\usepackage{units}

\usepackage{gensymb}

\usepackage[mode=buildnew]{standalone}

\usepackage{placeins}

\usepackage{import}

\usepackage[%
    margin=0pt,
]{caption}
\usepackage[list=true]{subcaption}

\usepackage{mathtools}
\DeclarePairedDelimiter\ceil{\lceil}{\rceil}
\DeclarePairedDelimiter\floor{\lfloor}{\rfloor}

\usepackage{amsmath,amssymb}
\DeclareMathOperator*{\minimize}{min}

\usepackage{xpatch}

\newlength{\tightalgowidth}
\newlength{\tightalgoremainder}

\newenvironment{tightalgo}[2][]
{%
	\setlength{\tightalgowidth}{#2}
	\setlength{\tightalgoremainder}{\linewidth-\tightalgowidth}
	\makeatletter
	\makeatother
	\begin{algorithm}[#1] %
}%
{ \end{algorithm} }

\makeatletter
\renewcommand{\algocf@makecaption@ruled}[2]{%
	\global\sbox\algocf@capbox{\hskip\AlCapHSkip
		\parbox[t]{\linewidth}{\algocf@captiontext{#1}{#2}}}
}%

\renewcommand{\algocf@caption@ruled}{\box\algocf@capbox\kern\interspacetitleruled\hrule  height\algotitleheightrule width \linewidth depth0pt\kern\interspacealgoruled}
\def\@algocf@pre@ruled{\hrule width \linewidth height \algoheightrule depth0pt\kern\interspacetitleruled}%
\def\@algocf@post@ruled{\kern\interspacealgoruled\hrule width \linewidth height\algoheightrule\relax}%
\patchcmd{\@algocf@start}{%
	\begin{lrbox}{\algocf@algobox}%
	}{%
		\begin{lrbox}{\algocf@algobox}%
			\begin{minipage}{\tightalgowidth}%
			}{}{}
			\patchcmd{\@algocf@finish}{%
			\end{lrbox}%
		}{%
		\end{minipage}%
	\end{lrbox}%
}{}{}
\makeatother

\definecolor{butter1}{rgb}{0.988,0.914,0.310}
\definecolor{chocolate1}{rgb}{0.914,0.725,0.431}
\definecolor{chameleon1}{rgb}{0.541,0.886,0.204}
\definecolor{skyblue1}{rgb}{0.447,0.624,0.812}
\definecolor{plum1}{rgb}{0.678,0.498,0.659}

\definecolor{bblue}{HTML}{4F81BD}
\definecolor{rred}{HTML}{C0504D}
\definecolor{ggreen}{HTML}{9BBB59}
\definecolor{ppurple}{HTML}{9F4C7C}

\definecolor{ablue}{HTML}{106e8d}
\definecolor{ared}{HTML}{c23a3f}
\definecolor{agreen}{HTML}{454c2c}

\pgfplotscreateplotcyclelist{bblue to rred}{%
	color=bblue, mark=diamond*\\%
	color=rred!12.5!bblue, mark=square*\\%
	color=rred!25!bblue, mark=triangle*\\%
	color=rred!37.5!bblue,mark=asterisk\\%
	color=rred!50!bblue,mark=otimes*\\%
	color=rred!62.5!bblue,mark=pentagon*\\%
	color=rred!75!bblue, mark=|\\%
	color=rred!87.5!bblue,mark=x\\%
	color=rred,mark=-\\%
}

\pgfplotscreateplotcyclelist{blue to red}{%
	color=blue, mark=diamond*\\%
	color=red!10!blue, mark=square*\\%
	color=red!20!blue, mark=triangle*\\%
	color=red!30!blue,mark=pentagon*\\%
	color=red!40!blue,mark=otimes*\\%
	color=red!50!blue,mark=asterisk\\%
	color=red!60!blue, mark=|\\%
	color=red!70!blue,mark=x\\%
	color=red!80!blue,mark=-\\%
	color=red!90!blue,mark=asterisk\\%
	color=red, mark=|\\%
}

\pgfplotsset{
	/pgfplots/bar legend image/.append style={
		legend image code/.code={
			\draw [##1] (0cm,-0.4ex) rectangle +(1.8em,1.2ex);
		}
	}
}

\newcommand*{\tikzmk}[1]{\tikz[remember picture,overlay,] \node (#1) {};\ignorespaces}

\let\year\ieeeaccessyear

\begin{document}

\title{Dataflow Aware Mapping of Convolutional Neural Networks Onto Many-Core Platforms With Network-on-Chip Interconnect}

\author{Andreas~Bytyn,
  René~Ahlsdorf,
  Rainer~Leupers,
  and~Gerd~Ascheid
  \thanks{A. Bytyn, R. Ahlsdorf, R. Leupers, and G. Ascheid are with the Institute of Communication Technologies and Embedded Systems (ICE), RWTH Aachen University, 52074 Aachen,
    Germany, e-mail: \{bytyn,ahlsdorf,leupers,ascheid\}@ice.rwth-aachen.de.}
  \thanks{This work was supported by the German Federal Ministry of Education and Research (BMBF) via the PARIS project (16ES0602).}}

\maketitle

\begin{abstract}
  Machine intelligence, especially using convolutional neural networks (CNNs), has become a large area of research over the past years. Increasingly sophisticated hardware accelerators are proposed that exploit e.g. the sparsity in computations and make use of reduced precision arithmetic to scale down the energy consumption. However, future platforms require more than just energy efficiency: Scalability is becoming an increasingly important factor. The required effort for physical implementation grows with the size of the accelerator making it more difficult to meet target constraints. Using many-core platforms consisting of several homogeneous cores can alleviate the aforementioned limitations with regard to physical implementation at the expense of an increased dataflow mapping effort. While the dataflow in CNNs is deterministic and can therefore be optimized offline, the problem of finding a suitable scheme that minimizes both runtime and off-chip memory accesses is a challenging task which becomes even more complex if an interconnect system is involved.
  This work presents an automated mapping strategy starting at the single-core level with different optimization targets for minimal runtime and minimal off-chip memory accesses. The strategy is then extended towards a suitable many-core mapping scheme and evaluated using a scalable system-level simulation with a network-on-chip interconnect. Design space exploration is performed by mapping the well-known CNNs AlexNet and VGG-16 to platforms of different core counts and computational power per core in order to investigate the trade-offs. Our mapping strategy and system setup is scaled starting from the single core level up to 128 cores, thereby showing the limits of the selected approach.
\end{abstract}


\section{Introduction}
Convolutional neural networks (CNNs) are nowadays widely used for applications such as face recognition and object detection. While smaller networks like MobileNet \cite{Howard2017} and SqueezeNet \cite{Iandola2016} can be efficiently processed using small dedicated accelerators or even ARM based general-purpose CPUs \cite{Ignatov}, larger networks require much more powerful dedicated hardware. In data-center applications, often GPUs are the method of choice due to their easy programmability and massive performance which, however, is paid for by their large power consumption compared to application-specific accelerators \cite{Cavigelli2017}. For future platforms, it is therefore desirable to have scalable performance with reasonable power consumption so that the accelerator can be tailored towards a certain set of problems.

Using a single large core yields high throughput and energy efficiency if designed for networks with specific dimensions. However, having a single very large accelerator makes it more difficult to find suitable mappings for CNNs with dimensionalities differing significantly from the initial design goals. As a result, the arithmetic units are often under-utilized as shown in \cite{Shen}. This issue can be circumvented by having many smaller cores that allow a more fine-grained mapping of the available data-slices. These cores should provide sufficient flexibility, e.g. by being programmable, in order to allow for multiple processing schemes to be mapped onto them. One possible way of achieving this is to use application-specific instruction-set processors (ASIPs) as shown in \cite{Bytyn2019}.

The choice of arithmetic (e.g. limited precision fixed-point, binary/log quantized etc.) has a large impact on the power consumption of a system, so does the dataflow scheme. In this work, the term \textit{dataflow} is used to describe both the temporal as well as the spatial distribution of data-packets, i.e. filter weights and feature maps of convolutional kernels, in the overall system. By maximizing the spatial correlation of data and minimizing the temporal dependencies, it is possible to reduce the number of off-chip accesses and, therefore, reduce the energy required for data movement. In the context of a many-core system it is, however, not sufficient to only optimize the dataflow for a single core but the optimization must be done for the entire system.

For many-core systems, a very important aspect is the choice of interconnect: While traditional crossbar based interconnects imply a moderate implementation complexity, they quickly become a bottleneck in terms of system scalability. Using a robust and well-proven network-on-chip (NoC), on the other hand, results in a larger initial implementation complexity that is rewarded by a much easier system scalability. For these reasons, we focus on configurable many-core systems that use a DRAM-centric NoC with a mesh topology as an interconnect.

While the dataflow of a single core can be accurately classified and evaluated based on its specific loop-order, loop-tiling and related loop-unrolling parameters \cite{Ma2017, Chen2017}, a sophisticated system-level simulation is required to do so for many-core systems due to the interconnect. The main reason for this is that congestion phenomena in the NoC are not easily predictable and must therefore be simulated. This is because the NoC requests issued by all the cores affect each other, i.e. causing stalls within some cores while waiting for data. The main contributions of this work can be summarized as follows:
\begin{itemize}
\item Based on previous work from \cite{Ma2017}, we provide a mathematical formulation for finding the optimal single-core mapping scheme for an existing programmable CNN accelerator with user-definable optimization targets: minimal runtime or minimal off-chip accesses.
\item The single-core mapping problem is extended towards the many-core case and a heuristic for finding automatic mappings is described.
\item Case studies for mappings of the well-known CNNs AlexNet and VGG-16 are presented and quantitatively evaluated using system-level simulations of a platform comprised of multiple accelerators and a network-on-chip.
\end{itemize}
The remainder of this work is structured as follows: In Section \ref{sec:state_of_the_art}, a brief overview of existing CNN accelerators and dataflow optimization schemes is given. Next, the simulation setup is elaborated in Section \ref{sec:simulation_framework}, whereas the implementation details of the different system components are presented in Section \ref{sec:system_components}. A solution for the single-core mapping problem is mathematically formulated in Section \ref{sec:single_core_mapping} and simulation results are depicted in Section \ref{sec:single_core_eval}. Our proposed mapping heuristic for extension towards the many-core case is detailed in Section \ref{sec:multi_core_flow}. Comprehensive simulation results for mappings onto different platform configurations are then presented in Section \ref{sec:multi_core_eval}. Section \ref{sec:conclusion} concludes this work with some final remarks.

\section{Background}
\label{sec:state_of_the_art}
\subsection{Hardware Acceleration of CNNs}
\label{subsec:hardware_acc_of_cnns}
Starting with the widespread use of CNNs in the early 2010's, there have been a number of hardware accelerators presented in both the research community as well as the industry. In general, the different types of accelerators can be grouped into one of the following categories: GPUs, FPGAs, ASICs and application-specific instruction-set processors (ASIPs), in which the latter two groups vary in terms of their data processing scheme. Some well-known examples of dedicated accelerators are Snowflake \cite{Gokhale2017} (FPGA), Escher \cite{Shen2017} (FPGA), Origami \cite{Cavigelli2017} (ASIC), Eyeriss \cite{Chen2016} (ASIC), Envision \cite{Moons2017} (ASIP) and ConvAix \cite{Bytyn2019} (ASIP). While they differ in terms of their specific processing scheme, they are all optimized towards the same overall goal: Keep data as local as possible to reduce off-chip transfers and maximize processing intensity, thereby maximizing the utilization of their arithmetic units. A key difference in the presented architectures is their degree of flexibility, ranging from fixed dataflow to runtime configurable dataflow. Of course, having a more flexible processing scheme always comes at some cost in terms of area (additional control units) and energy (setting up the processing scheme and multiplexing the data), but it also gives designers more degrees of freedom. This freedom can then be exploited when optimizing the mapping which is especially important for many-core platforms. For this reason, we propose the use of a flexible dataflow accelerator, e.g. an ASIP, and incorporate a suitable system-level model of such a core into our overall simulation setup as described in Section \ref{sec:simulation_framework}.

\subsection{Convolutional Layer}
\label{subsec:conv_layer}
Todays CNNs consist of a number of different layers, amongst them convolutional layers, pooling layers and activation layers. Due to the fact that the largest computational demand is generated by the regular convolutional layers, we focus our investigations on them. The convolutional operation can be described according to \eqref{eq:convop} as follows:
\begin{align}
&\boldsymbol{O}(c_o, y_o, x_o) = \boldsymbol{B}({c_o})\ + \nonumber \\
&\sum_{{c_i}=0}^{N_{if}} \sum_{{k_y}=0}^{N_{k_y}} \sum_{{k_x}=0}^{N_{k_x}} \boldsymbol{W}(c_o, c_i, k_y, k_x) \cdot \boldsymbol{I}(c_i, y_o \cdot s, x_o \cdot s)
\label{eq:convop}
\end{align}
whereby $\boldsymbol{O}$ depicts the entirety of all output feature maps (ofmaps) indexed by their channel $c_o$ and position $(y_o, x_o)$ within the ofmap, while $\boldsymbol{I}$ represents the set of all input feature maps (ifmaps) with the same indexing scheme as the ofmaps. The respective biases for each output channel are depicted by $\boldsymbol{B}$, with $\boldsymbol{W}$ being the weights associated with the ofmap channel $c_o$, ifmap channel $c_i$ and filter kernel position $(k_y, k_x)$. Lastly, $s$ represents the stride of the convolution. The limits of the sums represent the total number of ifmaps $N_{if}$ as well as the filter kernel's height $N_{ky}$ and width $N_{kx}$.
Since the result of \eqref{eq:convop} only calculates one single output pixel for one output channel, an even greater number of multiply-accumulate (MAC) operations is required to calculate all ofmaps completely. The actual implementation of this can be represented by many nested for-loops as shown in Algorithm~ \ref{alg:nested_for_loops}.
%
\addtolength{\skiprule}{-1.5mm}
\begin{tightalgo}[htb]{\linewidth}
    \LinesNumbered
    \KwIn{ifmaps $\boldsymbol{I}$, filter weights $\boldsymbol{W}$, biases $\boldsymbol{B}$, stride $s$}
    \KwResult{ofmaps $\boldsymbol{O}$}
    \For{$c_o = 0;\ c_o < N_{of};\ c_o\texttt{++}$}{
      \For{$y_o = 0;\ y_o < N_{oy};\ y_o\texttt{++}$}{
        \For{$x_o = 0;\ x_o < N_{ox};\ x_o\texttt{++}$}{
          $\boldsymbol{O}(c_o, y_o, x_o) = \boldsymbol{B}(c_o)$
          
          \For{$c_i = 0;\ c_i < N_{if};\ c_i\texttt{++}$}{
            \For{$k_y = 0;\ k_y < N_{ky};\ k_y\texttt{++}$}{
              \For{$k_x = 0;\ k_x < N_{kx};\ k_x\texttt{++}$}{
                $\boldsymbol{O}(c_o, y_o, x_o) \mathrel{+}= \boldsymbol{W}(c_o, c_i, k_y, k_x) \cdot \boldsymbol{I}(c_i, y_o \cdot s, x_o \cdot s)$
              }
            }
          }
        }
      }
    }
    \caption{Nested for-loops of a convolutional layer.}
    \label{alg:nested_for_loops}
\end{tightalgo}
\addtolength{\skiprule}{1.5mm}

\subsection{Dataflow Optimization}
\label{subsec:dataflow_optimization}

Related work on the topic of dataflow optimization can be split twofold: First into work related to the topic of mapping tasks onto multi/many-core systems in general and second into techniques focusing on the optimal slicing and tiling of CNNs on the single accelerator level.

For the first topic, a comprehensive overview is presented in \cite{Singh2013} where the authors introduce a taxonomy that allows to classify mapping techniques. A distinction between run-time and design-time mapping techniques is made and further differentiation is provided based on the type of target architecture, which can either be a heterogeneous or a homogeneous many-core system. For highly deterministic tasks like CNNs, the authors of \cite{Singh2013} suggest that using design-time mapping has the benefit of being able to optimize the overall system instead of having to rely on a more narrow view of the system. Several optimization goals such as performance, communication cost, energy consumption and reliability are elaborated, whereat in this work the focus is on the first three. The authors of \cite{Rouhani2017} focus on the CNN-specific task-mapping within a heterogeneous many-core system, e.g. consisting of multiple CPUs and FPGAs or GPUs. Their aim is to speed up the training of CNNs by optimizing the platform mapping, which is done via a depth-first graph traversal methodology that takes into account inter-core and inter-memory communication overhead.

Regarding the second topic, which is the optimized slicing and tiling of a CNN for a specific accelerator, several authors have proposed their own taxonomies and strategies. The authors of \cite{Chen2017} provide a taxonomy for CNN accelerators according to how data stationarity is realized, e.g. \textit{output stationary} in case the output feature map's pixels are kept in local scratchpad memories or registers. In \cite{Ma2017}, an elaborate set of design parameters is provided to mathematically describe this stationarity and subsequently the parameters are explored to find optimal configurations for given CNNs. These parameters, together with the order of loops as depicted in Algorithm \ref{alg:nested_for_loops}, fully describe how data is organized within an accelerator and, therefore, they are used in this work as well. For convenience, the most important design parameters are summarized in \mbox{Table \ref{tab:conv_params}}. While the first column of parameters describes the CNN layer's dimensions, the second column describes the loop tiling applied to the loops in Algorithm \ref{alg:nested_for_loops}. The last column expresses the loop unrolling factors which result in parallel computation being executed at each loop iteration. Often, this is equivalent to the fixed hardware parallelism within an accelerator and must be set at design-time. For example, unrolling the output width dimension $N_{ox}$ by setting $P_{ox}=N_{ox}$ would result in $P_{ox}$ MAC units working in parallel on a single channel's output row. In this work, a processing core with design-time configurable $P_{of}$ and $P_{ox}$ values is used as described in Section \ref{sec:system_components}.

\begin{table}[t]
  \centering
  \caption{Loop parameters for convolutional layers as proposed in \cite{Ma2017}.\vspace{-0.3cm}}
    \begin{tabular}{|c|r|c|c|c|}
\cline{3-5}    \multicolumn{1}{r}{} &       & Dimension & Tiling & Unrolling \bigstrut\\
    \hline
    \multirow{2}[4]{*}{Filter kernels} & Height & $N_{ky}$   & $T_{ky}$   & $P_{ky}$ \bigstrut\\
\cline{2-5}          & Width & $N_{kx}$   & $T_{kx}$   & $P_{kx}$ \bigstrut\\
    \hline
    \multirow{3}[6]{*}{IFMaps} & Height & $N_{iy}$   & $T_{iy}$   & $P_{iy}$ \bigstrut\\
\cline{2-5}          & Width & $N_{ix}$   & $T_{ix}$   & $P_{ix}$ \bigstrut\\
\cline{2-5}          & Channels & $N_{if}$   & $T_{if}$   & $P_{if}$ \bigstrut\\
    \hline
    \multirow{3}[6]{*}{OFMaps} & Height & $N_{oy}$   & $T_{oy}$   & $P_{oy}$ \bigstrut\\
\cline{2-5}          & Width & $N_{ox}$   & $T_{ox}$   & $P_{ox}$ \bigstrut\\
\cline{2-5}          & Channels & $N_{of}$   & $T_{of}$   & $P_{of}$ \bigstrut\\
    \hline
    \end{tabular}%
  \label{tab:conv_params}%
\end{table}%

The main focus of this work is the optimization of the tiling parameters as they are runtime configurable and largely determine the dataflow. It should be noted that the term \textit{mapping} in this work refers to a concrete set of aforementioned parameters with some additional slicing parameters which are related to the many-core case as introduced later. Since a detailed evaluation of all possible different loop orders is not possible within the scope of this work, we focus on the one presented in Algorithm~\ref{alg:nested_for_loops} with some modifications as described in Section \ref{sec:multi_core_flow} to cope with the many-core mapping.
%
%
\section{Simulation Setup}
\label{sec:simulation_framework}

To investigate the fitness of a CNN mapping in terms of runtime and communication cost, a simulation is required that is capable of accurately modeling both the single-core transactions as well as the actual communication via an interconnect network. While evaluation of a single-core mapping is possible based on analytical considerations only as shown in Section \ref{sec:single_core_mapping}, the many-core case requires to account for the communication overhead induced by the NoC, i.e. network congestion caused e.g. by limitations in data buffers. The most accurate results for such an analysis could be obtained by using a full RTL-level simulation of the NoC and the accelerator cores, however this would result in an unreasonably high runtime. We therefore use a parameterizable system-level simulation
in which different components of the system are implemented using approximately-timed transaction-level modeling (TLM) techniques as described in SystemC TLM \cite{SystemC_TLM}. Since the focus of this work is the investigation of the dataflow, each processing core is modeled in an abstract fashion: An inner process is defined that imitates the dataflow of the actual core in the way an external observer would see it. This is done by traversing the loop structure as previously described without actually performing any computations. However, the formation of data-packets (both send- and receive-packets) is carried out accurately and these packets are injected into the NoC at the corresponding times. Using a cycle-accurate instruction-set simulator of the processing core, we verified the correctness of the generated transactions.
All remaining components, e.g. the NoC router, are modeled in a cycle-accurate fashion in order to accurately reflect any congestion phenomena that might occur during execution. Furthermore, to allow realistic modeling of a complex system, the simulation supports two clock domains, one for the NoC running at a higher frequency and one for the processing cores. Also, sophisticated monitoring and tracing facilities are included, thereby allowing quantitative evaluation of e.g. the number of data-packets routed over a certain router, the number of SRAM and DRAM memory accesses performed, the count of MAC operations executed per core and the average port buffer stalling times in the NoC.


\subsection{System Overview}
\label{sec:system_components}
\begin{figure}[t]
	\centering
	\includegraphics[width=0.7\linewidth]{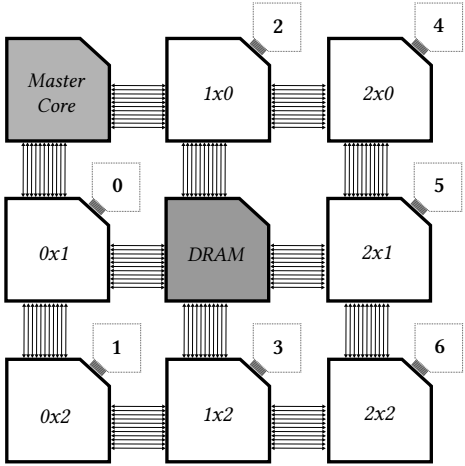}
	\caption{Example architecture of a 3x3 NoC with 7 processing cores, 1 master core and a DRAM interface block.}
	\label{fig:3x3_sys_architecture}
\end{figure}

As mentioned before, a 2D mesh-style NoC that uses a credit-based flow control with the XY routing scheme was selected as system interconnect. Our NoC implementation is based on the work presented in \cite{Moraes2004} with some adaptions made especially to the direct memory network interface (DMNI) \cite{Ruaro2016} which is responsible for managing the data flow between the processing cores and the NoC. One possible configuration of the interconnect is shown in Fig. \ref{fig:3x3_sys_architecture}. Each router has 4 ports for each direction and an additional local port connected to the processing element.

In the following sections, different mesh- and processing core sizes are investigated. However, since the base components always stay the same, they are briefly introduced here. An exhaustive investigation of different NoC parameters such as the flit width, packet length, port buffer sizes and the positioning of the DRAM interface were conducted for this work. Based on this investigation, we use a flit width of \unit[64]{bit} and a packet length of 40 flits per packet. The inport buffer size of the router is set to \unit[16]{flits} with the DRAM interface placed in the center of the mesh. If the mesh-size is increased beyond the 3x3 configuration depicted in Fig. \ref{fig:3x3_sys_architecture}, the DRAM block is always re-centered and the master core remains at position $(0,0)$ (top left). The additional positions within the mesh-grid are then filled with processing cores. Furthermore, the system uses two clock domains, one for the processing cores that runs at 500 MHz and one for the NoC that runs at a higher frequency of 1 GHz. According to \cite{DDRNoC}, this is a reasonable choice for implementing a NoC in a modern CMOS technology.

\subsection{Processing Core Architecture}
\label{subsec:processing_core}
In order for the mapper to generate the best possible results, it is desirable to have a processing core with the highest possible flexibility with regards to the dataflow. We therefore decided to use an application-specific instruction-set processor (ASIP) that is very similar to the one presented in \cite{Bytyn2019}.  The processor uses very long instruction words (VLIW) with 8 slots in parallel and a RISC-like instruction set architecture (ISA) that also incorporates some more complex vector instructions specifically targeted at CNNs.
%

The total number of parallel MAC operations that can be scheduled in one cycle depends on the design-time configurable unrolling factors $P_{ox} \in \lbrace 4,8,16,32 \rbrace$, $P_{of} \in \lbrace 4,8,16 \rbrace$ and is calculated as the product of these two factors. At the inner-most loop level according to Algorithm \ref{alg:nested_for_loops}, the ASIP uses an output row stationary scheme, i.e. one ofmap row of width $P_{ox}$ for $P_{of}$ ofmap channels is kept in the register file in parallel. Each MAC operation uses \unit[16]{bit} fixed-point multiplier operands that are accumulated in \unit[32]{bit} registers. Data for both weights and ifmaps is provided by a sophisticated on-chip memory interface and a direct memory access (DMA) controller that moves data between the on-chip SRAM and the external memory via the NoC. The size of the SRAM scales with the vector parallelism ($P_{ox}$) of the core and is calculated as follows: $D_{sram} = P_{ox} \cdot \unit[4096]{words} \cdot W_{word}$ with $W_{word}=\unit[16]{bit}$ being the wordwidth. We synthesized and placed and routed the ASIP using a 28nm TSMC technology node for standard operating conditions (\unit[1]{V}, \unit[25]{\degree}) resulting in a maximum clock frequency of \unit[500]{MHz} (\unit[400]{MHz} for the largest core with $P_{ox}=32$).
%
%
%
\subsection{Network-on-Chip Components}
The relevant components of the NoC used in this work are the network router, the direct memory access network interface (DMANI), the DRAM interface handling data accesses to the external DRAM memory and the master core that schedules computations onto the processing cores described in Section \ref{subsec:processing_core}. Our implementation is closely based on the work of the HERMES infrastructure \cite{Moraes2004} with some adaptions as described later on. Each packet within the NoC includes a flit containing the payload size, a header flit with a destination and - in contrast to HERMES - also a source address which is used by the DRAM interface as the destination for sending back data fetched due to a DRAM load request. More information on the separate components is given below.

\begin{figure}[t]
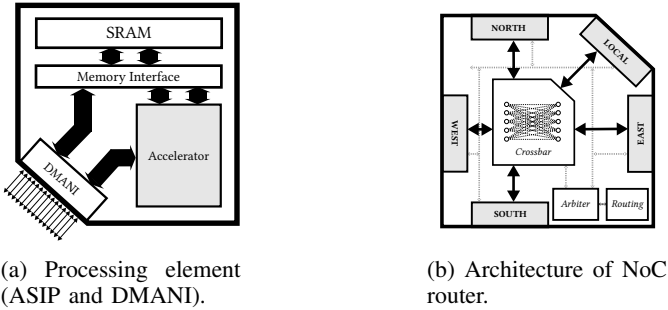

  \centering
  \subcaptionbox[]{%
    Processing element (ASIP and DMANI).%
    \label{subfig:noc_processing_entity}%
  }
  {%
    \includegraphics[width=0.36\linewidth]{figures/processing_entity.ai}%
  }%
  \hfill
  \subcaptionbox[]{%
    Architecture of NoC router.%
    \label{subfig:noc_router_design}%
  }
  {%
    \includegraphics[width=0.36\linewidth]{figures/router_design.ai}%
  }%
  \vspace{-0.2cm}
  \caption[]{NoC processing elements.}
  \label{fig:noc_elements}
\end{figure}

\textbf{DMANI:} The DMANI used in this work is an extension of the DMNI introduced in \cite{Ruaro2016} whose main task is to offload the NoC packet-handling from the processing core so that said core can focus on performing computations. While the original DMNI required the accelerator to set up every NoC packet, containing information such as the target address and payload size, our DMANI does this on its own. It is used on top of the already existing direct memory access (DMA) controller contained within the core which is responsible for keeping tabs on outstanding read and write transactions. Whenever the core issues a new transaction, it is handed over to the DMANI which then determines the number of required packets and returns a request ID to the DMA. All requests handed to the DMANI are processed in a FIFO fashion. To further reduce the overhead for the processing core, the DMANI has direct access to the core's SRAM memory via the core's memory interface as shown in Fig. \ref{subfig:noc_processing_entity}. So, in contrast to the original DMNI, no interrupt routine is required for the DMANI to write or read data to/from the SRAM. Instead, whenever a request is handed to the DMA of the core, it is ensured in software that a sufficiently large space within the SRAM is reserved. Access to the SRAM is arbitrated via the core's own memory interface which leaves the possibility of in-accessibility during an on-going transaction. To reduce the effects of such conflicts, the DMANI has a receive buffer for packets being received from the NoC (as response to an earlier issued DRAM read request) and a write-buffer for pre-fetching data in case of a write to the NoC. Data contained within a service packet is used to configure the processing cores at the beginning.

\textbf{Router:} For any NoC, the router is one of the main components that enables communication between the different processing entities. As mentioned before, our router design is based on the original HERMES router presented in \cite{Moraes2004} which has four bi-directional ports for connecting to other routers (\texttt{North}, \texttt{East}, \texttt{South}, \texttt{West}) and a local port that is connected to its processing core as depicted in Fig. \ref{subfig:noc_router_design}. The router consists of the following sub-components: crossbar, arbiter, routing module and individual port buffers. Each individual port buffer is responsible for requesting arbitration based on its stored requests. Arbitration is then handled according to a prioritization scheme that is as follows: \texttt{East}, \texttt{West}, \texttt{North}, \texttt{South}. Afterwards, the list is shifted in a cyclical fashion so that no starvation of requests occurs. A new packet is routed via the routing module by determining its destination address based on the XY routing protocol and the connection between the ports is established using the crossbar network. In total, this process, starting at the port buffers up to when the crossbar connection is established, takes 4 clock cycles.

\textbf{DRAM interface:} To allow access to an external DRAM memory, one of the grid-spaces within the NoC mesh is reserved for a DRAM interface that receives both read- and write-requests from the entire NoC and stores them in an internal request buffer that can save exactly one request per processing element. The interface preferably serves write-requests in order to minimize the backlog into the NoC that can occur in case of a long write request arising together with multiple other write requests. In general, incoming requests are processed in a FIFO fashion without any further prioritization. For this work, we assume a DRAM bus width equal to the NoC's flit width resulting in a maximum bandwidth of \unit[8]{GByte/s} at a NoC clock frequency of \unit[1]{GHz} and a flit width of \unit[64]{bit}, which is very reasonable even for a slow DDR memory.

\textbf{Master core:} The processing cores must be set up according to the mapping described in Section \ref{sec:multi_core_flow}. In this work, a master-slave concept is employed in which a master core, that could be a RISC-like microprocessor, assigns the configurations to the different cores via a service message using the already introduced NoC packet structure. Since the actual configuration is determined offline, the master core only has to send these pre-calculated configuration packets and is, therefore, modeled as a simple state machine.

To summarize, the main system parameters used in this work are depicted in Table \ref{tab:noc_parameters}.
\begin{table}[t]
  \centering
  \caption{System parameter overview.}
    \begin{tabular}{lr}
    \textbf{Parameter} & \textbf{Value} \bigstrut[b]\\
    \hline
    Max. packet length & \unit[40]{flits} \bigstrut[t]\\
    Flit width ($W_{flit}$) & \unit[64]{bit} \\
    NoC clock frequency ($f_{noc}$) & \unit[1]{GHz} \\
    Core clock frequency ($f_{core}$) & \unit[500]{MHz} \\
    Router inport buffer size & \unit[16]{flits} \\
    DMANI buffer size & \unit[64]{words} \\
    DRAM bandwidth ($\text{BW}_{dram}$) & $\unit[64]{\frac{bit}{cycle}}$
    \end{tabular}%
  \label{tab:noc_parameters}%
\end{table}

\subsection{Energy Modelling}
Since power and energy consumption are important metrics to determine the fitness of a mapping, we use a macro-modeling approach that estimates energy consumption at a high level. As mentioned before, relevant key figures with regards to energy consumption such as SRAM/DRAM load/store counts, number of MACs etc. are already traced in our simulation. These figures are used to estimate the overall energy consumption for the processing cores and DRAM according to \eqref{eq:energy_core} and \eqref{eq:energy_dram} as follows:
\begin{align}
E_{core}\ =\ &E_{idle} \cdot N_{cyc} + E_{mac} \cdot N_{mac}\ +\nonumber \\
&E_{sram\_ld} \cdot N_{sram\_ld}\ +\nonumber \\
&E_{sram\_st} \cdot N_{sram\_st} %
\label{eq:energy_core} \\
E_{dram}\ =\ &E_{dram\_ld} \cdot N_{dram\_ld}\ +\nonumber \\
&E_{dram\_st} \cdot N_{dram\_st} %
\label{eq:energy_dram}
\end{align}
where $E_x$ represents the energy value associated with a single event of type $x$ and $N_x$ represents the event-count during the whole simulation time. The energy values for the processing core were extracted from time-annotated post-layout simulation of a design similar to the one presented in \cite{Bytyn2019}. More details on the processing core are shown in Section \ref{subsec:processing_core}. Because these energy values were obtained using statistical methods, they already contain the energy required for e.g. program control of the processor (included in the idle energy) and register file accesses in case of MAC and memory operations. For DRAM memory accesses, we use an energy of \unit[21]{pJ/Bit} as reported by \cite{Schaffner2015} for LPDDR3 memory.

For estimating the NoC's energy consumption, we used the model presented in \cite{Chan2005} which associates energies with the most energy intensive events in NoC routers: Routing a packet ($E_{route}$), arbitrating a request ($E_{arb}$), setting up the crossbar ($E_{xbar\_su}$), switching of the crossbar ($E_{xbar\_sw}$), buffering an incoming packet ($E_{buf}$) and leakage ($E_{leak}$). The energy values used in this work are summarized in Table \ref{tab:energy_values}. Since the original values presented in \cite{Chan2005} were extracted from a \unit[90]{nm} CMOS technology which is different from the TSMC \unit[28]{nm} technology used for the core, all values were scaled to \unit[28]{nm} according to $E_{new}=E_{old} \left(\frac{V_{new}}{V_{old}}\right)^{2} \frac{N_{new}}{N_{old}}$ with $N$ being the gate pitch and $V$ being the supply voltage.
\begin{table}[htbp]
	\centering
	\begin{threeparttable}[b]
		\caption{Energy values used in the macro-model.}
		\label{tab:energy_values}%
		\begin{tabular}{@{}p{0.95\linewidth}@{}}
			\centering
			\begin{tabular}{lr|lr}
				\multicolumn{2}{c|}{\textbf{Processing Core \& DRAM}} & \multicolumn{2}{c}{\textbf{Network-on-Chip}} \bigstrut[b] \\
				Name  & \multicolumn{1}{l|}{Energy} & Name & \multicolumn{1}{l}{Energy} \bigstrut[b] \\
				\hline
				$E_{idle}$ & 148.42 pJ/Cycle & $E_{route}$ & 0.06 pJ/Packet \bigstrut[t] \\
				$E_{sram\_ld}$ & 0.89 pJ/Bit & $E_{arb}$ & 0.22 pJ/Packet \\
				$E_{sram\_st}$ & 0.46 pJ/Bit & $E_{xbar\_sw}$ & 0.03 pJ/Bit \\
				$E_{mac}$ & 6.42 pJ/Op \tnote{a} & $E_{xbar\_su}$ & 0.16 pJ/Bit \\
				$E_{dram\_ld}$ & 21 pJ/Bit & $E_{buf}$ & 0.09 pJ/Bit \\
				$E_{dram\_st}$ & 21 pJ/Bit & $E_{leak}$ & 0.43 pJ/Cycle
			\end{tabular}%
		\end{tabular}
		
		\begin{minipage}{\linewidth}
			\begin{tablenotes}
				\item [a] Energy for 1 MAC with 16-bit multiplier operands and 32-bit accumulator that both use saturating fixed-point arithmetic.
			\end{tablenotes}
		\end{minipage}
	\end{threeparttable}
\end{table}%
\section{Single-Core Dataflow Mapping}
\label{sec:single_core_mapping}

In order to find a suitable many-core mapping scheme, we use a bottom-up flow in which the single-core mapping, i.e. the determination of suitable tiling factors adhering to the taxonomy in Table~\ref{tab:conv_params}, is calculated first. The communication effects of the NoC are not considered in this step, thereby making it possible to use the high-level dataflow description from Algorithm \ref{alg:nested_for_loops} and formulate the mapping problem as a constrained mixed integer non-linear problem (MINLP). We introduce two different optimization targets: minimum overall computing time and minimum off-chip memory accesses. In the following, a derivation for the aforementioned cost-functions is given which then allows us to find optimal single-core tiling parameters using a regular MINLP solver.

Using knowledge of the actual software implementation of the convolutional layer as presented in Section \ref{subsec:conv_layer} on our programmable ASIP, the nested loop structure with tiling can be elaborated as shown in Algorithm \ref{alg:asip_loop}. Note that for brevity, the unrolled for-loops according to the selected $P_{ox}$ and $P_{of}$ values are not shown. For this work, we selected the following tiling dimensions as this made the most sense given current CNN topologies and the given hardware: tiling amongst the ofmap \& ifmap channels ($T'_{of}$, $T'_{if}$) and the ifmap width ($T'_{ix}$) which of course results in tiling along the ofmap width $T'_{ox}=(T'_{ix} - N_{kx})/s + 1$ (padding is already included in the ifmap width $T'_{ix}$). Furthermore, for examination of the single-core case, we use dashed values for the tile-size $T'_x$, tile-count $S'_x$ and dimension $N'_x$ ($x \in \lbrace of, if, ox, ix \rbrace$) values in order to differentiate these single-core optimization parameters from the later to be introduced many-core optimization parameters $T_x$ and $S_x$. The numbers of resulting tiles per dimension as denoted by $S'$ are calculated according to
\newcommand{\boxouter}[1]{\tikz[remember picture,overlay]{\path let \p1=(B) in coordinate (B) at (9cm,\y1 + 0.5mm);\node[rounded corners, yshift=3pt,fill=#1,opacity=.12,fit={($(A) + (0.3pt, 2pt)$)($(B)+(-4cm,.8\baselineskip)$)}] {};}\ignorespaces}
\newcommand{\boxinner}[1]{\tikz[remember picture,overlay]{\path let \p1=(B) in coordinate (B) at (8cm,\y1);\node[yshift=3pt,fill=#1,opacity=.12,fit={($(A) + (-0.1mm, 0.88mm)$)($(B)+(-7.7cm,.8\baselineskip - 4mm)$)},rounded corners] {};}\ignorespaces}
\addtolength{\skiprule}{-2.0mm}
\newcommand{\algoforindent}{\mbox{}\phantom{for (}}
\begin{tightalgo}[t]{2\linewidth}
  \LinesNumbered
  \KwIn{CNN paramters: $N_{kx/ky}, N_{if}, N'_{of}\ \text{etc.}, \text{stride}\ s$ \newline Tiling parameters: $T'_{of/if/ox}, S'_{of/if/ox}, P_{ox}, P_{of}$}
  \KwResult{Associated costs: $N_{sram\_ld/st}, N_{dram\_ld/st}, N_{mac}$}
  \tikzmk{A}
  \For{$t_o = 0;\ t_o < S'_{of};\ t_o\texttt{++}$}{\label{alg:asip_loop:l_sof}
    \For{$t_i = 0;\ t_i < S'_{if};\ t_i\texttt{++}$}{\label{alg:asip_loop:l_sif}
      $\rightarrow$ DMA\_Load\_Filters() \label{alg:asip_loop:dma_ld_filter} \\
      $\rightarrow$ DMA\_Load\_Biases() \label{alg:asip_loop:dma_ld_bias} \\
      \For{$t_x = 0;\ t_x < S'_{ox};\ t_x\texttt{++}$}{\label{alg:asip_loop:l_sox}
        $\rightarrow$ DMA\_Load\_IFMap\_Initial() \label{alg:asip_loop:dma_ld_ifmap_init} \\
        $\rightarrow$ DMA\_Load\_PSum\_Initial() \label{alg:asip_loop:dma_ld_psum_init} \\
        \tikzmk{B}
        \boxouter{white!10!red}
        \tikzmk{A}
        \For{$y_o = 0;\ y_o < N_{oy};\ y_o\texttt{++}$}{
          $\rightarrow$ DMA\_Load\_IFMap\_Next() \label{alg:asip_loop:dma_ld_ifmap_next} \\
          $\rightarrow$ DMA\_Load\_PSum\_Next() \label{alg:asip_loop:dma_ld_psum_next} \\
          \For{$c_o = t_o \cdot T'_{of};\ c_o < (t_o + 1) \cdot T'_{of};$ \\
            \algoforindent$c_o\texttt{+=}P_{of}$}{\label{alg:asip_loop_inner_start}
            \For{$x_o = t_x \cdot T'_{ox};\ x_o < (t_x + 1) \cdot T'_{ox};$ \\
              \algoforindent$x_o\texttt{+=}P_{ox}$}{
              $\rightarrow$ SRAM\_Load\_Bias\_or\_PSum() \label{alg:asip_loop:sram_ld} \\
              \For{$k_y = 0;\ k_y < N_{ky}; k_y\texttt{++}$}{\label{alg:asip_loop_innerst_start}
                \For{$c_i = t_i \cdot T'_{if};\ c_i < (t_i + 1) \cdot T'_{if};$ \\
                  \algoforindent$c_i\texttt{++}$}{
                  $\rightarrow$ Line\_Prefetch() \label{alg:asip_loop:prefetch} \\
                  \For{$k_x = 0;\ k_x < N_{kx};\ k_x\texttt{++}$}{
                    $\rightarrow$ MAC($P_{ox} \cdot P_{of}$ per cycle); \label{alg:asip_loop:mac}
                  }
                }
              }
              $\rightarrow$ SRAM\_Store\_OFMap\_or\_PSum()  \label{alg:asip_loop:sram_st}
            }
          }
          $\rightarrow$ DMA\_Store\_OFMap\_or\_PSum\_Row() %
          \tikzmk{B}%
          \label{alg:asip_loop:dma_st_ofmap} %
          \boxinner{white!10!blue}%
        } %
      } %
    } %
  } %
  \caption{Convolutional layer with tiling parameters as\newline implemented on the ASIP.}%
  \label{alg:asip_loop}%
\end{tightalgo}
\addtolength{\skiprule}{2.0mm}
\begin{align}
S'_{of} &= \ceil*{N'_{of}/T'_{of}} \label{eq:sof_singlecore} \\
S'_{if} &= \ceil*{N_{if}/T'_{if}} \label{eq:sif_singlecore} \\
S'_{ox} &= \ceil*{N'_{ox}/T'_{ox}} \label{eq:sox_singlecore}\text{.}
\end{align}
By annotating processing-cycle costs, e.g. for data prefetching and computations as well as data transfer-costs inferred by off-chip accesses, an overall cost for a tiling can be determined.
\begin{figure*}[tb]
	\centering
	\begin{minipage}{0.98\linewidth}
\begin{tikzpicture}
\pgfplotsset{
	xtick={0,1,...,12},
	 xticklabels={1\_1,1\_2,2\_1,2\_2,3\_1,3\_2,3\_3,4\_1,4\_2,4\_3,5\_1,5\_2,5\_3},
	x tick label style={anchor=north,font=\bfseries},
	legend style={
		legend columns=1,
		font=\footnotesize,
		at={(0.91\linewidth, -1.3cm)},
		anchor=east,
	},
}
\begin{axis}[thick,
height=4.0cm,
ymin=0,
ymax=40,
width=0.72\linewidth, xmin=0,xmax=12, axis y line*=right, axis x line=none, ylabel=Transfers (MByte),]%
\pgfplotsset{every outer y axis line/.style={xshift=1.63cm, dashed, very thick}, every tick/.style={xshift=1.63cm}, every y tick label/.style={xshift=1.63cm} }
\path [name path=origin] (0,0)
-- +(axis cs: 12,0);
\addplot[draw=none,name path=mincomp+dram] table[x index=0,y index=5, col sep=semicolon, /pgf/number format/read comma as period] {data/min_dram_comp__vgg16.csv};
\addplot[draw=none,name path=mindram+dram] table[x index=0,y index=6, col sep=semicolon, /pgf/number format/read comma as period] {data/min_dram_comp__vgg16.csv};
\addplot [white!10!bblue, opacity=0.3] fill between [
of=origin and mincomp+dram,
];
\addplot [black!20!rred, opacity=0.3] fill between [
of=origin and mindram+dram,
];

\end{axis} 

\begin{axis}[
thick,
title=VGG-16,
title style={
	font=\bfseries,
},
axis y line*=left,
ybar,
xlabel=Layer,
height=4.0cm,
width=0.72\linewidth,
ylabel=Runtime (ms),
ylabel style={yshift=1mm},
ymajorgrids,
bar width=2mm,
xmin=0,
ymin=0,
ymax=50,
ytick={0,10,20,...,50},
enlarge y limits=0,
enlarge x limits=0.05,
yminorgrids,
xtick pos = left,
scatter/position=absolute,
node near coords style={
	at={(axis cs:\pgfkeysvalueof{/data point/x} ,0)},
	anchor=east,
	rotate=65,
	yshift=-0.25cm,
	xshift=-0.15cm,
},
cycle list={
	{draw=rred, fill=rred},
	{draw=bblue,fill=bblue},
},
]
\addplot+[
draw=black, thick] table[y index=2, col sep=semicolon, /pgf/number format/read comma as period] {data/min_dram_comp__vgg16.csv};
\label{vggbarplotdram}
\addplot+[
draw=black, thick] table[y index=1, col sep=semicolon, /pgf/number format/read comma as period] {data/min_dram_comp__vgg16.csv};
\label{vggbarplotcomp}
\end{axis}

\begin{axis}[
thick,
axis y line*=left,
height=4.0cm,
width=0.72\linewidth,
xmin=0,
ymin=0,
ymax=25,
ytick={0,5,10,...,25},
enlarge y limits=0,
enlarge x limits=0.05,
cycle list={
	{draw=rred, fill=rred},
	{draw=bblue,fill=bblue},
},
axis y line*=right,
axis x line=none,
ymin=0,
ylabel=Energy (mJ),
ylabel style={yshift=-0.5mm},
]

\addlegendentry{\hspace{-.6cm}\textbf{Runtime}}\addlegendimage{empty legend}
\addlegendimage{bar legend image, black, fill=rred, thick}\addlegendentry{min-dram}
\addlegendimage{bar legend image, black, fill=bblue, thick}\addlegendentry{min-comp}
\addlegendentry{\hspace{-.6cm}\textbf{Energy}}\addlegendimage{empty legend}
\addlegendentry{min-dram}\addlegendimage{very thick, color=rred, mark=triangle*}
\addlegendentry{min-comp}\addlegendimage{very thick, color=bblue, mark=*}
\addlegendentry{\hspace{-.6cm}\textbf{Transfers}}\addlegendimage{empty legend}
\addlegendentry{min-dram}\addlegendimage{bar legend image, draw=black!20!rred, fill=black!20!rred, thin}
\addlegendentry{min-comp}\addlegendimage{bar legend image, fill=white!10!bblue, draw=white!10!bblue, thin}

\addplot[mark=*,bblue,very thick]
table[x index=0,y index=3, col sep=semicolon, /pgf/number format/read comma as period] {data/min_dram_comp__vgg16.csv};
\addplot[mark=triangle*,rred,very thick]
table[x index=0,y index=4, col sep=semicolon, /pgf/number format/read comma as period] {data/min_dram_comp__vgg16.csv};
\end{axis}
\end{tikzpicture}
\vspace{-2.0cm}
\end{minipage} %

\begin{minipage}{0.98\linewidth}
	\begin{tikzpicture}
	\pgfplotsset{
		xtick={0,1,...,7},
		xticklabels={1,2\_1,2\_2,3,4\_1,4\_2,5\_1,5\_2},
		x tick label style={anchor=north,font=\bfseries},
	}
	\begin{axis}[thick,
	height=4.0cm,
    xlabel=Layer,
	ymin=0,
	ymax=4,
	width=0.72\linewidth, xmin=0,xmax=7, axis y line*=right, axis x line=none, ylabel=Transfers (MByte),]%
	\pgfplotsset{every outer y axis line/.style={xshift=1.63cm, dashed, very thick}, every tick/.style={xshift=1.63cm}, every y tick label/.style={xshift=1.63cm} }
	\path [name path=origin] (0,0)
	-- +(axis cs: 7,0);
	\addplot[draw=none,name path=mincomp+dram] table[x index=0,y index=5, col sep=semicolon, /pgf/number format/read comma as period] {data/min_dram_comp__alexnet.csv};
	\addplot[draw=none,name path=mindram+dram] table[x index=0,y index=6, col sep=semicolon, /pgf/number format/read comma as period] {data/min_dram_comp__alexnet.csv};
	\addplot [white!10!bblue, opacity=0.3] fill between [
	of=origin and mincomp+dram,
	];
	\addplot [black!20!rred, opacity=0.3] fill between [
	of=origin and mindram+dram,
	];
	
	\end{axis} 
	
	\begin{axis}[
	thick,
	title=AlexNet,
	title style={
		font=\bfseries,
	},
	axis y line*=left,
	ybar,
	height=4.0cm,
	xlabel=Layer,
	width=0.72\linewidth,
	ylabel=Runtime (ms),
	ylabel style={yshift=1mm},
	ymajorgrids,
	bar width=2mm,
	xmin=0,
	ymin=0,
	ymax=2,
	ytick={0,0.5,...,2},
	enlarge y limits=0,
	enlarge x limits=0.05,
	yminorgrids,
	xtick pos = left,
	scatter/position=absolute,
	node near coords style={
		at={(axis cs:\pgfkeysvalueof{/data point/x} ,0)},
		anchor=east,
		rotate=65,
		yshift=-0.25cm,
		xshift=-0.15cm,
	},
	cycle list={
		{draw=rred, fill=rred},
		{draw=bblue,fill=bblue},
	},
	]
	\addplot+[
	draw=black, semithick] table[y index=2, col sep=semicolon, /pgf/number format/read comma as period] {data/min_dram_comp__alexnet.csv};
	\addplot+[
	draw=black, semithick] table[y index=1, col sep=semicolon, /pgf/number format/read comma as period] {data/min_dram_comp__alexnet.csv};
	\end{axis}
	
	\begin{axis}[
	thick,
	axis y line*=left,
	height=4.0cm,
	width=0.72\linewidth,
	xmin=0,
	ymin=0,
	ymax=2,
	ytick={0,0.5,...,2},
	enlarge y limits=0,
	enlarge x limits=0.05,
	cycle list={
		{draw=rred, fill=rred},
		{draw=bblue,fill=bblue},
	},
	axis y line*=right,
	axis x line=none,
	ymin=0,
	ylabel=Energy (mJ),
	ylabel style={yshift=-0.5mm},
	]
	
	\addplot[mark=*,bblue,very thick]
	table[x index=0,y index=3, col sep=semicolon, /pgf/number format/read comma as period] {data/min_dram_comp__alexnet.csv};
	\addplot[mark=triangle*,rred,very thick]
	table[x index=0,y index=4, col sep=semicolon, /pgf/number format/read comma as period] {data/min_dram_comp__alexnet.csv};
	\end{axis}
	\end{tikzpicture}
\end{minipage}
\caption{Resulting single-core runtimes (bars), DRAM transfers (filled areas) and energy consumption (lines) of the CNNs VGG-16 (top) and AlexNet (bottom) for two different optimization targets: \textit{min-dram} and \textit{min-comp}.}
\label{fig:single_core_eval}
\end{figure*}
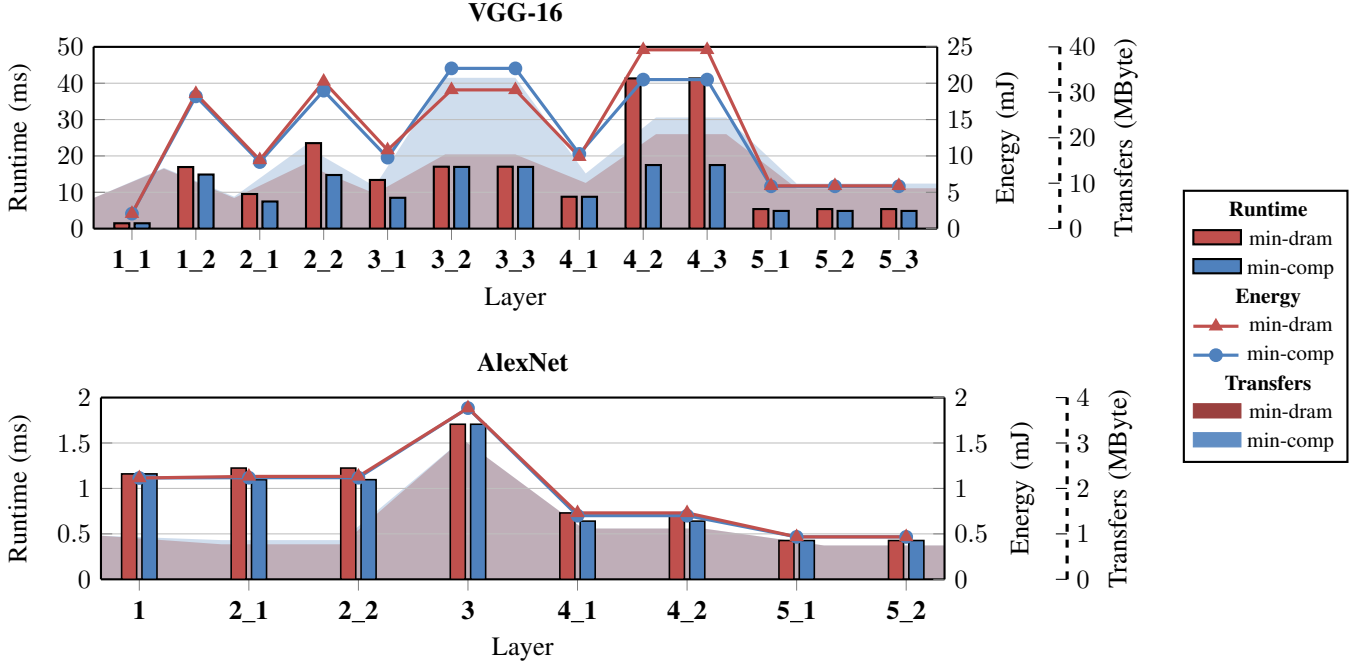
As depicted in Algorithm \ref{alg:asip_loop}, each step in the loop can be associated with a certain action, e.g. the DMA loading filters (line \ref{alg:asip_loop:dma_ld_filter}) or the calculation of a certain number of MAC operations (line \ref{alg:asip_loop:mac}). We subdivide these loops into an inner part that does not unconditionally depend on any off-chip accesses (marked in {\bfseries\color{bblue}blue}) and an outer part that directly relies on them (marked in {\bfseries\color{rred}red}). Based on this information, we first derive the total number of off-chip accesses $N_{dram} = N_{dram\_init} + N_{dram\_par}$. It is hereby important to differentiate between DMA requests that can be handled in parallel to the computations (line \ref{alg:asip_loop:dma_ld_ifmap_next}, \ref{alg:asip_loop:dma_ld_psum_next} and \ref{alg:asip_loop:dma_st_ofmap}), denoted by $N_{dram\_par}$, and those that must be waited for (line \ref{alg:asip_loop:dma_ld_filter}, \ref{alg:asip_loop:dma_ld_bias}, \ref{alg:asip_loop:dma_ld_ifmap_init} and \ref{alg:asip_loop:dma_ld_psum_init}), denoted by $N_{dram\_init}$.
\begin{align}
  N_{dram\_init} =&\ N'_{of} \cdot N_{kx} \cdot N_{ky} \cdot N_{if} && \text{\small(filters)} \nonumber \\
                 +&\ N'_{of} && \text{\small(biases)} \nonumber \\
                 +&\ S'_{of} \cdot N'_{ix} \cdot N_{ky} \cdot N_{if} && \text{\small(initial ifmaps)} \nonumber \\
                 +&\ (S'_{if}-1) \cdot N'_{ox} \cdot N'_{of} && \text{\small(initial psums)} \label{eq:dram_acc_init}
\end{align}%
\begin{align}
  &N_{dram\_par} = S'_{if} \cdot N'_{ox} \cdot N_{oy} \cdot N'_{of} && \text{\small(ofmap/psum store)} \nonumber \\
  &+\ S'_{of} \cdot N'_{ix} \cdot (N_{iy} - N_{ky}) \cdot N_{if} && \text{\small(next ifmaps)} \nonumber \\
  &+\ (S'_{if}-1) \cdot N'_{ox} \cdot (N_{oy}-1) \cdot N'_{of} && \text{\small(next psums)} \label{eq:dram_acc_next}
\end{align}
The required pure computational cycles within the inner loops (marked in {\bfseries\color{bblue}blue}) can be computed as follows:
\begin{align}
  C_{comp}   =&\ (C_{mac} + C_{sram}) \cdot N_{oy} && \text{} \label{eq:cycles_comp} \\
  C_{mac}    =&\ (C_{pfetch} + N_{kx}) \cdot T'_{if} \cdot N_{ky} \cdot \frac{T'_{ox}}{P_{ox}} \cdot \frac{T'_{of}}{P_{of}} && \text{}  \label{eq:cycles_mac} \\
  C_{pfetch} =&\ \ceil*{\frac{\text{Stride}+1}{2}}-1 && \text{}  \label{eq:cycles_prefetch} \\
  C_{sram}   =&\ 2 \cdot \frac{ T'_{ox} \cdot T'_{of} \cdot P_{ox} \cdot P_{of}}{ \text{BW}_{sram} } && \text{}  \label{eq:cycles_sram}
\end{align}
The prefetch cycle count $C_{pfetch}$ in \eqref{eq:cycles_prefetch} is specific to the processing core used in this work. The term $\text{BW}_{sram}$ represents the bandwidth in words per cycle that the internal SRAM offers and is equal to $2 \cdot P_{ox}$ here since the on-chip memory is a banked dual-port memory with bank count equal to $P_{ox}$. As mentioned before, certain DMA requests run in parallel to the computations, so to allow accurate calculation of the processing cycles for the inner loops, the estimated cycle count $C_{dram\_par}$ for these accesses must be calculated as well:
\begin{align}
  C_{dram\_par} =&\ \frac{N_{dram\_par}}{\text{BW}_{dram}} && \text{} \label{eq:cycles_dram_par}
\end{align}
in which the term $\text{BW}_{dram}$ represents the bandwidth in words per cycle that the DRAM can provide. In this work, aforementioned bandwidth is set to the NoC flit width per cycle divided by the wordwidth and multiplied with the clock ratio between the NoC and the core:
\begin{align}
\text{BW}_{dram} = \unit[64]{\frac{bit}{cycle}} / \unit[16]{\frac{bit}{word}} \cdot \frac{\unit[1]{GHz}}{\unit[500]{MHz}}= \unit[8]{\frac{word}{cycle}} \text{.}
\end{align}

The cycle count for the outer loops $C_{outer\_loop}$ depends solely on the time required to finish the initial DRAM accesses $N_{dram\_init}$.
\begin{align}
  C_{outer\_loop} =&\ \frac{N_{dram\_init}}{\text{BW}_{dram}} && \text{} \label{eq:cycle_outer_loop}
\end{align}
Depending on the selected tiling parameters, either the computational cycles $C_{comp}$ or the DRAM cycles $C_{dram\_par}$ will determine the overall inner loop cycles $C_{inner\_loop}$, which is modeled in our optimization problem as two inequality constraints according to \eqref{eq:constraint_inner_cycles_comp} and \eqref{eq:constraint_inner_cycles_dram}:
\begin{align}
  C_{inner\_loop} &\geq C_{comp} \cdot S'_{ox} \cdot S'_{if} \cdot S'_{of} \label{eq:constraint_inner_cycles_comp} \\
  C_{inner\_loop} &\geq C_{dram\_par} \label{eq:constraint_inner_cycles_dram}
\end{align}
Together with the previously calculated outer loop cycle count $C_{outer\_loop}$, the total cycle count $C_{total}$ amounts to:
\begin{align}
  C_{total} =&\ C_{outer\_loop} + C_{inner\_loop} && \text{} \label{eq:cycle_total}
\end{align}
Last but not least, the allocated SRAM memory that depends on the tiling parameters must be constrained since the individual processing cores only have limited on-chip memory available:
\begin{align}
  N_{sram\_alloc} =&\ \underbrace{T'_{of}}_{biases} + \underbrace{T'_{of} \cdot N_{kx} \cdot N_{ky} \cdot T'_{if}}_{filters} && \text{} \nonumber \\
                  +&\ \underbrace{T'_{if} \cdot (N_{ky} + stride) \cdot T'_{ix}}_{ifmaps} + \underbrace{3 \cdot T'_{ox} \cdot T'_{of}}_{ofmaps} && \text{} \label{eq:sram_alloc} \\
  N_{sram\_alloc} \leq&\ D_{sram} = P_{ox} \cdot \unit[8]{KByte} = P_{ox} \cdot \unit[4096]{word} && \text{} \label{eq:constraint_sram}
\end{align}
As can be seen in \eqref{eq:sram_alloc}, there are always 3 ofmap rows allocated because we use a triple-buffering scheme: One allocated row is used for pre-fetching the next partial sums (line \ref{alg:asip_loop:dma_ld_psum_next} in Algorithm \ref{alg:asip_loop}), one is used for calculating the current ofmap row and the third is used as write-back buffer (needed in line \ref{alg:asip_loop:dma_st_ofmap}).

Depending on the optimization target, it is now possible to find tiling parameters either minimizing the computational cycles (hereafter referred to as \textit{min-comp}) or the total number of DRAM accesses (referred to as \textit{min-dram}). The final optimization targets are shown in \eqref{eq:opt_min_comp} and \eqref{eq:opt_min_dram}.
\begin{align}
  \minimize_{T'_{of}, T'_{if}, T'_{ox}} C_{total} && \text{\small(\textit{min-comp})} \label{eq:opt_min_comp} \\
  \minimize_{T'_{of}, T'_{if}, T'_{ox}} N_{dram\_init} + N_{dram\_par} && \text{\small(\textit{min-dram})} \label{eq:opt_min_dram}
\end{align}
For brevity, the previously introduced constraints for the minimization targets are not restated here but must be added to the problem formulation.
\section{Single-Core Mapping Evaluation}
\label{sec:single_core_eval}

The previously introduced single-core mapping algorithm is used to map two well known CNNs, VGG-16 \cite{Simonyan2014} and AlexNet \cite{Krizhevsky2012}, onto a 3x1 system configuration with just one processing core, a master core and a DRAM interface block. Using our system-level simulation, the runtime, number of DRAM accesses as well as the energy consumption are obtained and the results are presented in Fig. \ref{fig:single_core_eval}. As can be seen, the \textit{min-comp} target always achieves faster runtime compared to the \textit{min-dram} target, however, this is achieved at the expense of an increased DRAM access count. Since AlexNet is a fairly small CNN compared to the much larger VGG-16, there are no large differences between both optimization targets. However, even though DRAM energy consumption is generally considered to be the main contributor in terms of overall energy, our results show that the total energy for VGG-16 is actually minimized for the \textit{min-comp} case. The causes for this observation are layers \textit{4\_2} and \textit{4\_3} which exhibit a much longer runtime when optimized for \textit{min-dram}. This can be explained as follows: To minimize the number of DRAM accesses, the optimization algorithm uses a configuration in which the width of the ofmap tiles $T'_{ox}$ is fairly small. In doing so, the on-chip SRAM is used to store a maximally large number of ifmap channels at the same time ($T'_{if}$ very large) thereby minimizing the need for any partial sum (psum) transfers. The small width of the ofmap tiles ($T'_{ox}$), however, causes a bad utilization of the processing core's vALUs ($T'_{ox} < P_{ox}$) leading to the increased runtime. Finally, the energy that is consumed for baseline processing core operation during this additional runtime results in the overall higher energy consumption.
\section{Many-Core Dataflow Mapping}
\label{sec:multi_core_flow}
To enable a large-scale evaluation of different CNNs onto arbitrary platform configurations, it is indispensable to have an automated method to assign computing tasks to the processing cores. The same taxonomy that was used for tiling the CNN layery in the single-core case (Section \ref{sec:single_core_mapping}) is, therefore, used here as well. While the dashed single-core parameters such as tile-size $T'_x$ and tile-count $S'_x$ referred to data dimensions within one core, the un-dashed variables have a different meaning. These variables, namely $T_x$ and $S_x$, are used to express the size and count of complete CNN layer slices in the many-core context. So starting from the top-level, each CNN layer is subdivided into a number of slices along the ifmap and ofmap width dimension ($S_{ix}$ and $S_{ox}$) as well as the ofmap channel dimension $S_{of}$ as depicted in Fig.~ \ref{subfig:slice_graphic}. For such a slice, it is then possible to determine an optimal single-core mapping as previously derived. An overview of the many-core mapping heuristic is given in Fig.~ \ref{fig:manycore_flowchart} with a more thorough explanation in the following paragraphs.

In order to determine a suitable mapping it is indispensable to first define a cost function. Since we always encounter either a computation bound or an I/O bound for our application, it makes the most sense to optimize for these targets by incorporating the overall runtime and time required for communication via the NoC into the cost function. The resulting optimization goal is depicted in the following:
\begin{align}
\underset{\mathbf{s}}{\text{min}}\ & \bigl\lparen \underset{\forall c \in\ \mathbb{C}}{\text{max}}(C_{tot\_wo\_dram}(\mathbf{s}_{c})) \nonumber \\
&+ \frac{1}{\text{BW}_{dram}} \cdot \sum_{\forall c \in\  \mathbb{C}} \sum_{p \ \in\ \mathbb{P}_{c}} F(p) \cdot W_{flit} \bigr\rparen
\label{eq:manycore_costfunc}
\end{align}
The first term in \eqref{eq:manycore_costfunc} represents among all processing cores $c \in \mathbb{C}$ the one with the maximum cycle count $C_{tot\_wo\_dram}(\mathbf{s}_{c})$ not counting any cycles required for DRAM accesses. An abstract mapping vector $\mathbf{s}_c$ hereby denotes the assignment of slices to processing cores. Taking the maximum is required here because in case of asymmetric mapping of slices to cores there might be a few cores left computing at the end, thereby dictating the runtime of the overall system. It can be calculated for each core as follows (see \eqref{eq:cycles_comp} and \eqref{eq:constraint_inner_cycles_comp} for reference):
\begin{align}
C_{tot\_wo\_dram} = C_{comp} \cdot S'_{ox} \cdot S'_{if} \cdot S'_{of}
\label{eq:cycle_tot_wo_dram}
\end{align}
In addition, the second term of \eqref{eq:manycore_costfunc} is used to model the overall NoC bandwidth requirement. To this end, the number of flits $F(p)$ for all NoC packets $p \in \mathbb{P}_{c}$ generated by each core $c$ is calculated. This is done by traversing the loop structure from Algorithm \ref{alg:asip_loop} and building an exact list of all packets with their associated lengths because only in doing so the overhead for e.g. having many small packets with associated header-flits can be accounted for.
\tikzset{%
    >={Latex[width=2mm,length=2mm]},
    base/.style = {rectangle, rounded corners, draw=black,
        text width=5.0cm, minimum height=1.0cm,
        text centered, font=\sffamily},
    detSlices/.style = {base, fill=blue!30},
    evalMap/.style   = {base, fill=red!30},
    storeMap/.style  = {base, fill=green!30},
    process/.style   = {base, minimum width=2.5cm, fill=orange!15},
    exit/.style      = {circle, x radius=2cm, y radius=1cm, fill=orange!30},
}
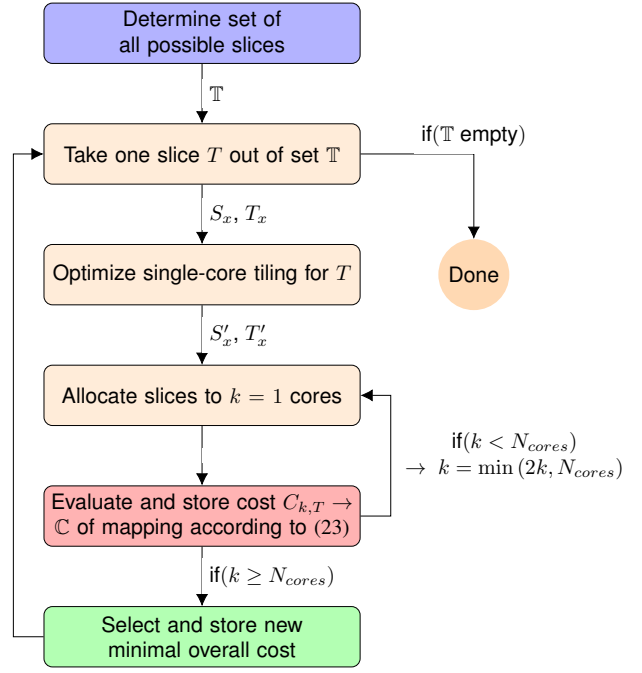
\begin{figure}[t]
 \centering
 \resizebox{0.95\linewidth}{!}{
    \begin{tikzpicture}[node distance=2.0cm,%
    every node/.style={fill=white, font=\sffamily},%
    align=center]
    \node (start)             [detSlices]                        {Determine set of all possible slices};
    \node (takeSlice)         [process, below of=start]          {Take one slice $T$ out of set $\mathbb{T}$};
    \node (optSlice)          [process, below of=takeSlice]      {Optimize single-core tiling for $T$};
    \node (waveAssign)        [process, below of=optSlice]       {Allocate slices to $k=1$ cores};
    \node (mapCost)           [evalMap, below of=waveAssign]     {Evaluate and store cost $C_{k,T} \rightarrow \mathbb{C}$ of
                                                                  mapping according to \eqref{eq:manycore_costfunc}};
    \node (storeMap)          [storeMap, below of=mapCost]       {Select and store new minimal overall cost};
    \node (exit)              [exit, right of=takeSlice,
                               xshift=2.5cm, yshift=-2.0cm]      {Done};
    
    \draw[->]     (start)        -- node[anchor=west] {$\mathbb{T}$}    (takeSlice);
    \draw[->]     (takeSlice)    -- node[anchor=west] {$S_x$, $T_x$}    (optSlice);
    \draw[->]     (optSlice)     -- node[anchor=west] {$S'_x$, $T'_x$}  (waveAssign);
    \draw[->]     (waveAssign)   -- node[anchor=west] {} (mapCost);
    \draw[->]     (mapCost.east) -- ++(0.5,0) -- ++(0,1) -- ++(0,1) --   
                                    node[xshift=2.3cm,
                                         yshift=-1.0cm]{$\text{if}(k < N_{cores})$ \\ $\rightarrow\ k=\min{(2k,N_{cores})}$} (waveAssign.east);
    \draw[->]     (mapCost)      -- node[anchor=west] {$\text{if}(k \ge N_{cores})$} (storeMap);
    \draw[->]     (storeMap.west) -- ++(-0.5,0) -- ++(0,8) -- ++(0,0) --  (takeSlice.west);
    \draw[->]     (takeSlice.east) -| node[anchor=south]{$\text{if}(\mathbb{T}\ \text{empty})$}  (exit.north);
    \end{tikzpicture}
  }
  \caption{Flow-chart of many-core mapping heuristic.}
  \label{fig:manycore_flowchart}
\end{figure}
\begin{figure*}[tb]
  \centering
  \hfill
  \subcaptionbox[]{%
    \label{subfig:slice_graphic}%
  }
  {%
    \centering
    \includegraphics[width=0.25\linewidth]{figures/slicing_example_vertical.ai} %
  }%
  \hfill
  \subcaptionbox[]{%
    \label{subfig:manycore_constant_analysis}%
  }%
  {%
    \centering
    \begin{tikzpicture}
    \pgfplotstableset{
      /pgf/number format/read comma as period,
    };
    \begin{axis}[
    thick,
    /pgfplots/every axis plot post/.append style={
      very thick,
    },
    ymode=log,
    log ticks with fixed point,
    height=4.3cm,
    width=0.72\linewidth,
    ylabel=Runtime (ms),
    ylabel style={yshift=-2mm},
    xlabel style={yshift=0.5mm},
    ymajorgrids,
    yminorgrids,
    xmajorgrids,
    every major grid/.style={black, opacity=0.5},
    every minor grid/.style={black, opacity=0.1},
    xlabel=Cores, 
    enlarge y limits=0.05,
    enlarge x limits=0.05,
    xtick pos = left,
    minor y tick num=4,
    xtick=data,
    symbolic x coords={2,4,16,32,64,128},
    cycle list name=bblue to rred,
    legend style={
      font=\small,
      at={(0.5,-0.32)},anchor=north, legend columns=6},
    ]
    \addlegendentry{\textbf{VGG16 layers:}}\addlegendimage{empty legend}
    \addplot+[] table[x index=0,y index=1, col sep=semicolon, skip coords between index={0}{1}] {data/mpsoc_constant_analysis.csv};
    \addlegendentry{1\_1};
    \addplot+[] table[x index=0,y index=2, col sep=semicolon, skip coords between index={0}{1}] {data/mpsoc_constant_analysis.csv};
    \addlegendentry{1\_2};
    \addplot+[] table[x index=0,y index=3, col sep=semicolon, skip coords between index={0}{1}] {data/mpsoc_constant_analysis.csv};
    \addlegendentry{2\_1};
    \addplot+[] table[x index=0,y index=4, col sep=semicolon, skip coords between index={0}{1}] {data/mpsoc_constant_analysis.csv};
    \addlegendentry{2\_2};
    \addplot+[] table[x index=0,y index=5, col sep=semicolon, skip coords between index={0}{1}] {data/mpsoc_constant_analysis.csv};
    \addlegendentry{3\_1};
    \addlegendentry{}\addlegendimage{empty legend}
    \addplot+[] table[x index=0,y index=6, col sep=semicolon, skip coords between index={0}{1}] {data/mpsoc_constant_analysis.csv};
    \addlegendentry{3\_\{2,3\}};
    \addplot+[] table[x index=0,y index=7, col sep=semicolon, skip coords between index={0}{1}] {data/mpsoc_constant_analysis.csv};
    \addlegendentry{4\_1};
    \addplot+[] table[x index=0,y index=8, col sep=semicolon, skip coords between index={0}{1}] {data/mpsoc_constant_analysis.csv};
    \addlegendentry{4\_\{2,3\}};
    \addplot+[] table[x index=0,y index=9, col sep=semicolon, skip coords between index={0}{1}] {data/mpsoc_constant_analysis.csv};
    \addlegendentry{5\_\{1,2,3\}};
    \fill[rred,opacity=0.2] (70,-50) rectangle (130,16);
    \draw[dotted, color=black, very thick] (70,-50) -- (70,16);
    \draw[dotted, color=black, very thick] (130,-50) -- (130,16);
    \end{axis}
    \end{tikzpicture}
  }%
  \hfill
  \vspace{-0.2cm}
  \caption[]{(\subref{subfig:slice_graphic}) Illustration of the many-core slicing. (\subref{subfig:manycore_constant_analysis}) Results for simulation of VGG-16 using a system-setup with constant overall computing capabilities and RAM, i.e. \mbox{$N_{cores}$ x $(P_{ox} \cdot P_{of})$} \& \mbox{$N_{cores}$ x $D_{sram}$} are constant (\mbox{total MAC per cycle = 2048}, \mbox{total SRAM = \unit[1]{MByte}}).}
  \label{fig:slice_and_manycore_analysis}
  \vspace{-0.2cm}
\end{figure*}
Since the calculation of each core's computation cycles according to \eqref{eq:cycle_tot_wo_dram} requires knowledge of the single-core tiling parameters $S'_{x}$, $T'_{x}$ which in turn can only be calculated based on the slicing parameters $S_x$ and $T_x$, we propose an iterative scheme as depicted in Fig.~ \ref{fig:manycore_flowchart} that uses a heuristic to determine the latter parameters. This iterative process starts with determining a set $\mathbb{T}$ of all possible slice parameters based on the constraint that the width $T_{ox}$ and depth $T_{of}$ of a slice should ideally be a multiple of the processing cores' unrolling factors $P_{ox}$ and $P_{of}$ to maximally utilize the cores:
\begin{align}
\mathbb{T}:= \bigl\{ (m P_{of}, n P_{ox})\ |\ \forall\ m \in \left[ 1, \floor*{\frac{N_{of}}{P_{of}}} \right]; \nonumber \\
\ \forall\  n\in \left[1, \floor*{\frac{N_{ox}}{P_{ox}}}\right] \bigr\}
\label{eq:mso_tile_set}
\end{align}
Afterwards, each possible slice parameter set $T = \left(T_{of}, T_{ox}\right) \in \mathbb{T}$ is fed into the single-core tiling optimizer from Section \ref{sec:single_core_mapping} and the set of optimal tiling parameters ($T'_x$, $S'_x$) is determined. Since each slice can be viewed as a new CNN layer of smaller dimension, the single-core optimizer is fed with the slice's dimensions as follows:
\begin{align}
N'_{ox} &= T_{ox} \\
N'_{ix} &= \left( T_{ox} - 1\right) \cdot s + N_{kx} \\
N'_{of} &= T_{of}
\end{align}
Due to the constraint in the selection of the slice parameters, it only takes a few seconds to determine all possible tiling parameters. For a given slice parameter set $T$, the number of resulting slices for each dimension is calculated as follows:
\begin{align}
S_{ox} &= \ceil*{N_{ox}/T_{ox}} \\
S_{of} &= \ceil*{N_{of}/T_{of}}
\end{align}
As our investigations have shown especially for small CNNs such as AlexNet, it is not always desirable to distribute tasks to all processing cores. In fact, the cost for additional communication required by a large number of cores (second term in \eqref{eq:manycore_costfunc}) can greatly outweigh the potential saving in computational time, thereby rendering a solution with more cores slower than a competing solution with fewer active cores. Also, in addition to being slower, more active cores result in more energy being consumed during idle cycles, which again is not desirable. We found that using a waving scheme which systematically explores configurations with different numbers of active cores, starting with the ones closest to the DRAM interface block, gives better results compared to using all cores all of the time for every layer. 

So in a final step, for each slice parameter set $T$ and associated tiling parameters $T'_x$ and $S'_x$, aforementioned waving scheme is employed to distribute slices to core instances in the NoC. For the first step, all slices are mapped to a single core ($k=1$). After this, the number of activated cores $k$ is doubled and all tasks are assigned to the 2 cores closest to the DRAM. Again, the number of activated cores is doubled and the former steps are repeated until a configuration with all cores activated was tested. For each of these iterations, the overall cost $C_{k,T}$ is saved in a set $\mathbb{C}$. Afterwards, the configuration corresponding to the lowest cost $\min\{\mathbb{C}\}$ is selected as final configuration. It should be noted that during slice assignment our algorithm makes sure to map slices with adjacent boundaries in the ofmap width dimension (neighboring $T'_{ox}$ slices) to the same core which are then stitched together afterwards to remove the need for redundant filter loads. This method has proven to be very robust and fairly fast: Determining a mapping usually only takes a few minutes for e.g. VGG-16.

%
%
%
\section{Many-Core Mapping Evaluation}
\label{sec:multi_core_eval}

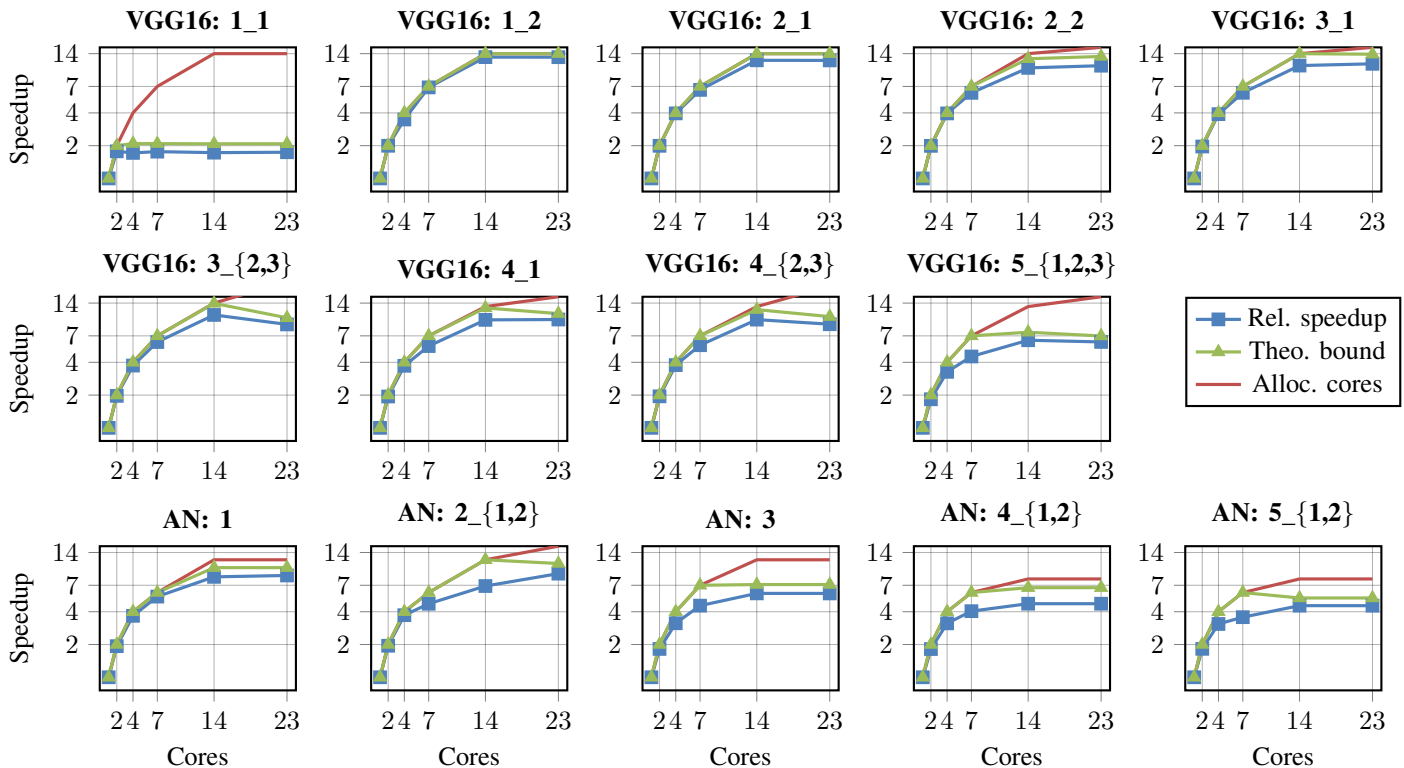
\begin{figure*}[htb]
  \begin{tikzpicture}
  \pgfplotstableset{
    /pgf/number format/read comma as period,
    col sep=semicolon,
  };
  \begin{groupplot}[
  group style={
    group name=Multi-core-plots,
    group size=5 by 3,
    ylabels at=edge left,
    vertical sep=1.4cm,
  },
  /pgfplots/group/every plot/.append style={
    thick,
    /pgfplots/every axis plot post/.append style={
      very thick,
    },
  },
  title style={
    yshift=-1mm,
  },
  height=3.5cm,
  width=0.231\textwidth,
  tickpos=left,
  ytick align=outside,
  xtick align=outside,
  enlarge x limits=0.05,
  ymode=log,
  log ticks with fixed point,
  xtick={2,4,7,14,23},
  ymin=0,
  ymax=16,
  ytick={2,4,7,14,23},
  minor x tick num=1,
  minor y tick num=1,
  scaled x ticks=false,
  grid=both,
  minor tick num=3,
  every major grid/.style={black, opacity=0.3},
  every minor grid/.style={black, opacity=0.1},
  ]
  
  \nextgroupplot[title={\textbf{VGG16: 1\_1}},ylabel=Speedup, ylabel style={yshift=0}]
  \addplot[thick, bblue, mark=square*] table[x index=0, y index=1] {data/mpsoc_multiple_sizes_speedups_vgg16.csv};
  \addlegendentry{Rel. speedup};
  \addplot[thick, rred, mark=circle*] table[x index=0, y index=1] {data/mpsoc_multiple_sizes_alloc_cores_vgg16.csv};
  \addlegendentry{Alloc. cores};
  \addplot[thick, ggreen, mark=triangle*] table[x index=0, y index=1] {data/mpsoc_multiple_sizes_theolim_vgg16.csv};
  \addlegendentry{Theo. bound};
  \legend{};
  
  \nextgroupplot[title={\textbf{VGG16: 1\_2}}]
  \addplot[thick,   bblue, mark=square*] table[x index=0, y index=2] {data/mpsoc_multiple_sizes_speedups_vgg16.csv};
  \addlegendentry{Rel. speedup};
  \addplot[thick, rred, mark=circle*] table[x index=0, y index=2] {data/mpsoc_multiple_sizes_alloc_cores_vgg16.csv};
  \addlegendentry{Alloc. cores};
  \addplot[thick, ggreen, mark=triangle*] table[x index=0, y index=2] {data/mpsoc_multiple_sizes_theolim_vgg16.csv};
  \addlegendentry{Theo. bound};
  \legend{};
  
  \nextgroupplot[title={\textbf{VGG16: 2\_1}}]
  \addplot[thick,   bblue, mark=square*] table[x index=0, y index=3] {data/mpsoc_multiple_sizes_speedups_vgg16.csv};
  \addlegendentry{Rel. speedup};
  \addplot[thick, rred, mark=circle*] table[x index=0, y index=3] {data/mpsoc_multiple_sizes_alloc_cores_vgg16.csv};
  \addlegendentry{Alloc. cores};
  \addplot[thick, ggreen, mark=triangle*] table[x index=0, y index=3] {data/mpsoc_multiple_sizes_theolim_vgg16.csv};
  \addlegendentry{Theo. bound};
  \legend{};
  
  \nextgroupplot[title={\textbf{VGG16: 2\_2}}]
  \addplot[thick,   bblue, mark=square*] table[x index=0, y index=4] {data/mpsoc_multiple_sizes_speedups_vgg16.csv};
  \addlegendentry{Rel. speedup};
  \addplot[thick, rred, mark=circle*] table[x index=0, y index=4] {data/mpsoc_multiple_sizes_alloc_cores_vgg16.csv};
  \addlegendentry{Alloc. cores};
  \addplot[thick, ggreen, mark=triangle*] table[x index=0, y index=4] {data/mpsoc_multiple_sizes_theolim_vgg16.csv};
  \addlegendentry{Theo. bound};
  \legend{};
  
  \nextgroupplot[title={\textbf{VGG16: 3\_1}}]
  \addplot[thick,   bblue, mark=square*] table[x index=0, y index=5] {data/mpsoc_multiple_sizes_speedups_vgg16.csv};
  \addlegendentry{Rel. speedup};
  \addplot[thick, rred, mark=circle*] table[x index=0, y index=5] {data/mpsoc_multiple_sizes_alloc_cores_vgg16.csv};
  \addlegendentry{Alloc. cores}
  \addplot[thick, ggreen, mark=triangle*] table[x index=0, y index=5] {data/mpsoc_multiple_sizes_theolim_vgg16.csv};
  \addlegendentry{Theo. bound}
  \legend{};
  
  \nextgroupplot[ylabel=Speedup, ylabel style={yshift=0mm},title={\textbf{VGG16: 3\_\{2,3\}}}]
  \addplot[thick,   bblue, mark=square*] table[x index=0, y index=6] {data/mpsoc_multiple_sizes_speedups_vgg16.csv};
  \addlegendentry{Rel. speedup};
  \addplot[thick, rred, mark=circle*] table[x index=0, y index=6] {data/mpsoc_multiple_sizes_alloc_cores_vgg16.csv};
  \addlegendentry{Alloc. cores};
  \addplot[thick, ggreen, mark=triangle*] table[x index=0, y index=6] {data/mpsoc_multiple_sizes_theolim_vgg16.csv};
  \addlegendentry{Theo. bound};
  \legend{};
  
  \nextgroupplot[title={\textbf{VGG16: 4\_1}}]
  \addplot[thick,   bblue, mark=square*] table[x index=0, y index=7] {data/mpsoc_multiple_sizes_speedups_vgg16.csv};
  \addlegendentry{Rel. speedup};
  \addplot[thick, rred, mark=circle*] table[x index=0, y index=7] {data/mpsoc_multiple_sizes_alloc_cores_vgg16.csv};
  \addlegendentry{Alloc. cores};
  \addplot[thick, ggreen, mark=triangle*] table[x index=0, y index=7] {data/mpsoc_multiple_sizes_theolim_vgg16.csv};
  \addlegendentry{Theo. bound};
  \legend{};
  
  \nextgroupplot[title={\textbf{VGG16: 4\_\{2,3\}}}]
  \addplot[thick,   bblue, mark=square*] table[x index=0, y index=8] {data/mpsoc_multiple_sizes_speedups_vgg16.csv};
  \addlegendentry{Rel. speedup};
  \addplot[thick, rred, mark=circle*] table[x index=0, y index=8] {data/mpsoc_multiple_sizes_alloc_cores_vgg16.csv};
  \addlegendentry{Alloc. cores};
  \addplot[thick, ggreen, mark=triangle*] table[x index=0, y index=8] {data/mpsoc_multiple_sizes_theolim_vgg16.csv};
  \addlegendentry{Theo. bound};
  \legend{};
  
  \nextgroupplot[title={\textbf{VGG16: 5\_\{1,2,3\}}}]
  \addplot[thick,   bblue, mark=square*] table[x index=0, y index=9] {data/mpsoc_multiple_sizes_speedups_vgg16.csv};
  \addlegendentry{Rel. speedup};
  \addplot[thick, rred, mark=circle*] table[x index=0, y index=9] {data/mpsoc_multiple_sizes_alloc_cores_vgg16.csv};
  \addlegendentry{Alloc. cores};
  \addplot[thick, ggreen, mark=triangle*] table[x index=0, y index=9] {data/mpsoc_multiple_sizes_theolim_vgg16.csv};
  \addlegendentry{Theo. bound};
  \legend{};
  
  \nextgroupplot[hide axis,legend style={at={(0,1)},
    anchor=north west,legend columns=1},]
  \addplot+[forget plot, draw=none,mark=none] coordinates {(1,1) (2,2)};
  \addlegendimage{mark=square*,bblue}
  \addlegendentry{Rel. speedup}
  \addlegendimage{mark=triangle*,ggreen}
  \addlegendentry{Theo. bound}
  \addlegendimage{mark=none, rred}
  \addlegendentry{Alloc. cores}
  
  \nextgroupplot[xlabel=Cores, xlabel style={yshift=-0.0mm},title={\textbf{AN: 1}},ylabel=Speedup, ylabel style={yshift=0}]
  \addplot[thick,   bblue, mark=square*] table[x index=0, y index=1] {data/mpsoc_multiple_sizes_speedups_alexnet.csv};
  \addlegendentry{Rel. speedup};
  \addplot[thick, rred, mark=circle*] table[x index=0, y index=1] {data/mpsoc_multiple_sizes_alloc_cores_alexnet.csv};
  \addlegendentry{Alloc. cores};
  \addplot[thick, ggreen, mark=triangle*] table[x index=0, y index=1] {data/mpsoc_multiple_sizes_theolim_alexnet.csv};
  \addlegendentry{Theo. bound};
  \legend{};
  
  \nextgroupplot[xlabel=Cores, xlabel style={yshift=-0.0mm},title={\textbf{AN: 2\_\{1,2\}}}]
  \addplot[thick,   bblue, mark=square*] table[x index=0, y index=2] {data/mpsoc_multiple_sizes_speedups_alexnet.csv};
  \addlegendentry{Rel. speedup};
  \addplot[thick, rred, mark=circle*] table[x index=0, y index=2] {data/mpsoc_multiple_sizes_alloc_cores_alexnet.csv};
  \addlegendentry{Alloc. cores};
  \addplot[thick, ggreen, mark=triangle*] table[x index=0, y index=2] {data/mpsoc_multiple_sizes_theolim_alexnet.csv};
  \addlegendentry{Theo. bound};
  \legend{};
  
  \nextgroupplot[xlabel=Cores, xlabel style={yshift=-0.0mm},title={\textbf{AN: 3}}]
  \addplot[thick,   bblue, mark=square*] table[x index=0, y index=3] {data/mpsoc_multiple_sizes_speedups_alexnet.csv};
  \addlegendentry{Rel. speedup};
  \addplot[thick, rred, mark=circle*] table[x index=0, y index=3] {data/mpsoc_multiple_sizes_alloc_cores_alexnet.csv};
  \addlegendentry{Alloc. cores};
  \addplot[thick, ggreen, mark=triangle*] table[x index=0, y index=3] {data/mpsoc_multiple_sizes_theolim_alexnet.csv};
  \addlegendentry{Theo. bound};
  \legend{};
  
  \nextgroupplot[xlabel=Cores, xlabel style={yshift=-0.0mm},title={\textbf{AN: 4\_\{1,2\}}}]
  \addplot[thick,   bblue, mark=square*] table[x index=0, y index=4] {data/mpsoc_multiple_sizes_speedups_alexnet.csv};
  \addlegendentry{Rel. speedup};
  \addplot[thick, rred, mark=circle*] table[x index=0, y index=4] {data/mpsoc_multiple_sizes_alloc_cores_alexnet.csv};
  \addlegendentry{Alloc. cores};
  \addplot[thick, ggreen, mark=triangle*] table[x index=0, y index=4] {data/mpsoc_multiple_sizes_theolim_alexnet.csv};
  \addlegendentry{Theo. bound};
  \legend{};
  
  \nextgroupplot[xlabel=Cores, xlabel style={yshift=-0.0mm},title={\textbf{AN: 5\_\{1,2\}}}]
  \addplot[thick,   bblue, mark=square*] table[x index=0, y index=5] {data/mpsoc_multiple_sizes_speedups_alexnet.csv};
  \addlegendentry{Rel. speedup};
  \addplot[thick, rred, mark=circle*] table[x index=0, y index=5] {data/mpsoc_multiple_sizes_alloc_cores_alexnet.csv};
  \addlegendentry{Alloc. cores};
  \addplot[thick, ggreen, mark=triangle*] table[x index=0, y index=5] {data/mpsoc_multiple_sizes_theolim_alexnet.csv};
  \addlegendentry{Theo. bound};
  \legend{};
  
  \end{groupplot}
  \end{tikzpicture}
  \caption{Speedups, number of allocated cores and theoretically bounded speedup according to \eqref{eq:theo_bound} for CNNs AlexNet (AN) and VGG-16 using systems with $N_{cores} \in \lbrace 2,4,7,14,23 \rbrace$ processing cores ($P_{ox}=16$, $P_{of}=8$, $D_{sram}=128 KByte$ per core). Layers with same dimensions and therefore same mappings and results are clustered as indicated by e.g. VGG16~ 5\_\{1,2,3\}.}
  \label{fig:manycore_results}
\end{figure*}

Using the previously described mapping scheme for the many-core case, a number of benchmarks are evaluated in this section. From a VLSI design perspective, it is much more appealing to build medium-sized processing cores and scale the system by adding more cores. This trend can be seen in the design choices made by manufacturers of large GPUs that usually consist of hundreds or even thousands of smaller RISC-like cores. Therefore, we simulate a system that employs a varying number of cores (4..128) which are scaled up or down (larger/smaller $P_{ox}$ and $P_{of}$ values) in such a way that the overall computing capabilities and on-chip SRAM of the system always stay at \unit[2048]{$\frac{\text{MAC}}{\text{cycle}}$} and \unit[1]{MByte} respectively. Looking at the results for running VGG-16 in Fig. \ref{subfig:manycore_constant_analysis}, it can be concluded that while having many small cores is not optimal due to the increased communication overhead in the NoC, also having only a few large cores does not give the best results. A medium-sized configuration with 16 processing cores (each core with configuration $P_{ox}=16$, $P_{of}=8$) achieves the fastest runtime for most layers as this yields the best trade-off between generating additional NoC communication on the one hand and enabling more degrees of freedom for the mapping algorithm on the other hand. It should be noted though that for networks with much larger ifmap widths, e.g. CNNs for semantic scene segmentation, it might be more favorable to include these larger cores as they provide a more natural match in terms of processing row width ($P_{ox}$). For these kinds of CNNs, however, the DRAM interface becomes the bottleneck no matter how optimal the mapping is.
To further explore the performance scalability of our approach, we select the most promising processing core configuration as determined earlier ($P_{ox}=16$, $P_{of}=8$) and simulate the same workload, i.e. the convolutional layers of VGG-16 and AlexNet, for a system configuration starting just with a single core going up to a 5x5 NoC configuration with 23 processing cores in total. The uneven core-counts of e.g. 23 for a 5x5 NoC can be explained by the 2 positions required for the DRAM interface and the master core. Calculating the overall computational capability of each configuration (measured in MAC per second) can easily be done as follows: $N_{MAC}=N_{cores} \cdot P_{ox} \cdot P_{of}$ (each core has $D_{sram} = \unit[128]{KByte}$ of SRAM). For the smallest system, this results in $N_{MAC}=128$ while the largest 23-core system offers up to $N_{MAC}=2944$. To obtain fair values for the following speed-up comparison of a many-core system compared to the single-core case, the single-core 3x1 configuration is simulated with a very large packet-length of $10000 \frac{\text{flit}}{\text{packet}}$. This is done in order to minimize any unnecessary communication overhead that would not be required in the single-core scenario. The results in Fig. \ref{fig:manycore_results} show the achieved speed-up relative to aforementioned single-core setup. Also, a line representing the theoretical bound is plotted which takes into consideration the tiling and slicing parameters generated by the algorithms introduced in Section \ref{sec:single_core_mapping} and Section \ref{sec:multi_core_flow} as well as the maximum available DRAM bandwidth. Said line denotes the speed-up that is achievable for given constraints without accounting for any of the overhead generated by the NoC and is calculated as follows:
\begin{align}
f(N_{cores}) = \frac{C_{single\_core}}{\max \left( C_{tot\_wo\_dram}, \frac{N_{dram}}{BW_{dram}} \right)}  \label{eq:theo_bound}
\end{align}
where $C_{single\_core}$ represents the number of processing cycles for the single-core setup.

The final simulation results in Fig. \ref{fig:manycore_results} show a close match between the theoretical bound as calculated by \eqref{eq:theo_bound} and the actual system performance whereas the explanation for the gap between both curves is twofold: On the one hand, the overhead required for NoC communication including congestion phenomena slightly diminishes the achievable speedup. On the other hand, the width and depth of the last slices generated according to \eqref{eq:mso_tile_set} are not necessarily a multiple of the cores' hardware parallelism $P_{ox}$ and $P_{of}$, resulting in a core under-utilization for these last slices. On average, the difference between the simulation results and the theoretical bound is between $6.59\%$ ($N_{cores}=2$) and $27.48\%$ ($N_{cores}=7$) for AlexNet and between $3.28\%$ ($N_{cores}=2$) and $17.32\%$ ($N_{cores}=14$) for VGG-16. Another interesting observation is the fact that the mapping heuristic does not choose to activate more than 14 cores for any of the configurations even if more cores are available. This can be explained by the large additional NoC traffic that would be generated and, as a result, would actually slow down the overall system. Looking at the curves of Fig.~ \ref{fig:manycore_results}, this trend of diminishing returns on investment can already be observed starting at 7 cores for most layers.

In general, most configurations with more than 7 cores are limited by the DRAM bandwidth with a few exceptions: Layer 1\_2 and 2\_1 of VGG-16 achieve very high speed-ups of up to 13x and 12.2x respectively for configurations with 14 cores. Since aforementioned layers are fairly wide ($N_{ix}=\lbrace 224, 112 \rbrace$) and only have a limited number of output channels ($N_{of}=\lbrace 64,128 \rbrace$), the mapping heuristic preferably slices along the width dimension. This minimizes the need for repeated loading of ifmaps thereby minimizing the bandwidth requirements and enabling the observed speedups. For later layers, this is not possible due to the larger ofmap channel counts and associated allocation of on-chip memory for filter-weights. This limitation could be overcome by enabling each core in the NoC to access the other cores' on-chip memories in order to reduce the traffic going through the DRAM interface and thereby using the cores' SRAM as caches.
\section{Conclusion}
\label{sec:conclusion}

Detailed strategies for mapping CNNs onto both single-core as well as many-core systems were presented and evaluated with regards to their practical feasibility. For the single-core mapping, two strategies, one for DRAM access minimization and one for runtime reduction, were investigated with the result that DRAM access minimization does not always lead to the desired energy reduction of the overall system. By using a system-level simulator, several system setups with different numbers of processing cores and core configurations were investigated. The results show that for the investigated CNNs AlexNet and VGG-16 configurations with 128 MAC units per core result in the highest speedups although this highly depends on the selected hardware unrolling factors $P_x$. Subsequent simulations that use this optimal core size have shown the speedup potential which saturates at a 4x4 system setup comprising 14 processing cores. The maximum speedups observed for this 4x4 system were 8.4x for AlexNet's first layer and 13x for VGG-16's second layer. Both layers share in common that they allow slicing in the ofmap width dimension which maps well to the given processing core's architecture. Since unrolling in the ofmap width and channel direction is very common among other accelerators as well, similar results could possibly be obtained for them too. As usual for many-core systems, ultimately the maximum achievable speed-up is always limited by the available on-chip memory and off-chip bandwidth. The use of a NoC for the investigated application domain has shown promising results and the presented mapping heuristic has proven to make robust choices for a variety of system configurations.

%
%
%
%
\bibliographystyle{IEEEtran}
\bibliography{IEEEabrv,mybib}
\end{document}